\documentclass[%
aip,
cha,
reprint
]{revtex4-2}

\usepackage[T1]{fontenc}

\usepackage{graphicx}

\usepackage{amssymb}
\usepackage{amsmath}
\usepackage{bm}

\usepackage{mathptmx}
\usepackage{etoolbox}

\makeatletter
\def\@email#1#2{%
 \endgroup
 \patchcmd{\titleblock@produce}
  {\frontmatter@RRAPformat}
  {\frontmatter@RRAPformat{\produce@RRAP{*#1\href{mailto:#2}{#2}}}\frontmatter@RRAPformat}
  {}{}
}%
\makeatother

\begin{document}
\title[Phase autoencoder for limit-cycle oscillators]{Phase autoencoder for limit-cycle oscillators}

\author{Koichiro Yawata}
\affiliation{Department of Systems and Control Engineering, Tokyo Institute of Technology, Tokyo 152-8552, Japan}
\email{koichiro.yawata.rt@gmail.com}
\author{Kai Fukami}
\affiliation{Department of Mechanical and Aerospace Engineering, University of California, Los Angeles, CA 90095, USA}
\author{Kunihiko Taira}
\affiliation{Department of Mechanical and Aerospace Engineering, University of California, Los Angeles, CA 90095, USA}
\author{Hiroya Nakao}
\affiliation{Department of Systems and Control Engineering, Tokyo Institute of Technology, Tokyo 152-8552, Japan}

\date{\today}

\begin{abstract}
We present a {\it phase autoencoder} that encodes the asymptotic phase of a limit-cycle oscillator, a fundamental quantity characterizing its synchronization dynamics.
This autoencoder is trained in such a way that its latent variables directly represent the asymptotic phase of the oscillator.
The trained autoencoder can perform two functions without relying on the mathematical model of the oscillator: 
first, it can evaluate the asymptotic phase and phase sensitivity function of the oscillator;
second, it can reconstruct the oscillator state on the limit cycle in the original space from the phase value as an input.
Using several examples of limit-cycle oscillators, we demonstrate that the asymptotic phase and phase sensitivity function can be estimated only from time-series data by the trained autoencoder.
We also present a simple method for globally synchronizing two oscillators as an application of the trained autoencoder.
\end{abstract}

\maketitle

\begin{quotation}
Spontaneous rhythmic phenomena are widely observed in the real world. They are generally modeled as limit-cycle oscillators in dynamical systems theory, and phase reduction theory has proven useful for understanding synchronization phenomena of interacting limit-cycle oscillators. In phase reduction theory, the state of the oscillator is represented by using a single variable called the asymptotic phase, but it is computationally challenging to evaluate; reproducing the original oscillator state from the phase value can also be difficult. In this study, we propose a machine-learning method based on the autoencoder to address these issues, which facilitates evaluation of the asymptotic phase only from time-series data and reproduction of the oscillator state from the phase value as an input. Our phase autoencoder can be used for data-driven synchronization control of limit-cycle oscillators without relying on their mathematical models.
\end{quotation}

\section{Introduction}\label{sec:intro}

Spontaneous rhythmic phenomena are widely observed in the real world and play important roles in the 
functioning of biological or engineered systems~\cite{Winfree1980,Pikovsky2001,Kuramoto2003,Stankovski2015,Stankovski2017,Borgius2014,Collins1993,Kobayashi2016,Funato2016,Kralemann2013,Garcia1998,rohden2012self}, such as heartbeats and respiration\cite{Kralemann2013}, brain waves\cite{Stankovski2015, Stankovski2017} and power grids~\cite{rohden2012self}.
Such rhythmic phenomena can be mathematically modeled 
as limit-cycle oscillators in nonlinear dynamical systems, and phase reduction theory~\cite{Winfree1980, Kuramoto2003, nakao2016phase,Ermentrout2010,monga2019phase,kuramoto2019concept,ermentrout2019recent} 
has been widely used for understanding the synchronization dynamics of coupled limit-cycle oscillators.

In phase reduction theory, the multidimensional state of a limit-cycle oscillator is described by a single phase variable
that increases with a constant natural frequency along the limit cycle and in its basin of attraction, called the asymptotic phase.
The phase sensitivity function (PSF, aka infinitesimal phase resetting curve) 
calculated from the asymptotic phase is important for characterizing the dynamics of weakly perturbed limit-cycle oscillators
and has been studied both theoretically and experimentally~\cite{Winfree1980, Ermentrout2010, glass1988clocks,nakao2016phase,shirasaka2020phase,wilson2022adaptive,taira2018phase}.

By phase reduction, the state of an oscillator can be approximately described by a simple phase equation
characterized only by the natural frequency and PSF of the oscillator, which is useful for analyzing
the synchronization dynamics of the oscillator in detail~\cite{Kuramoto2003,Hoppensteadt1997,Winfree1980,Ermentrout2010}.
Recently, more general phase-amplitude reduction theory has also been developed~\cite{monga2019phase,wedgwood2013phase,wilson2016isostable,mauroy2016global,shirasaka2017phase,mauroy2018global,kotani2020nonlinear,shirasaka2020phase,nakao2021phase,takata2021fast}, which is an extension of the phase reduction theory to include the deviations of the oscillator state from the limit cycle using exponentially decaying amplitudes.

\begin{figure}[htbp]
\includegraphics[scale=0.42]{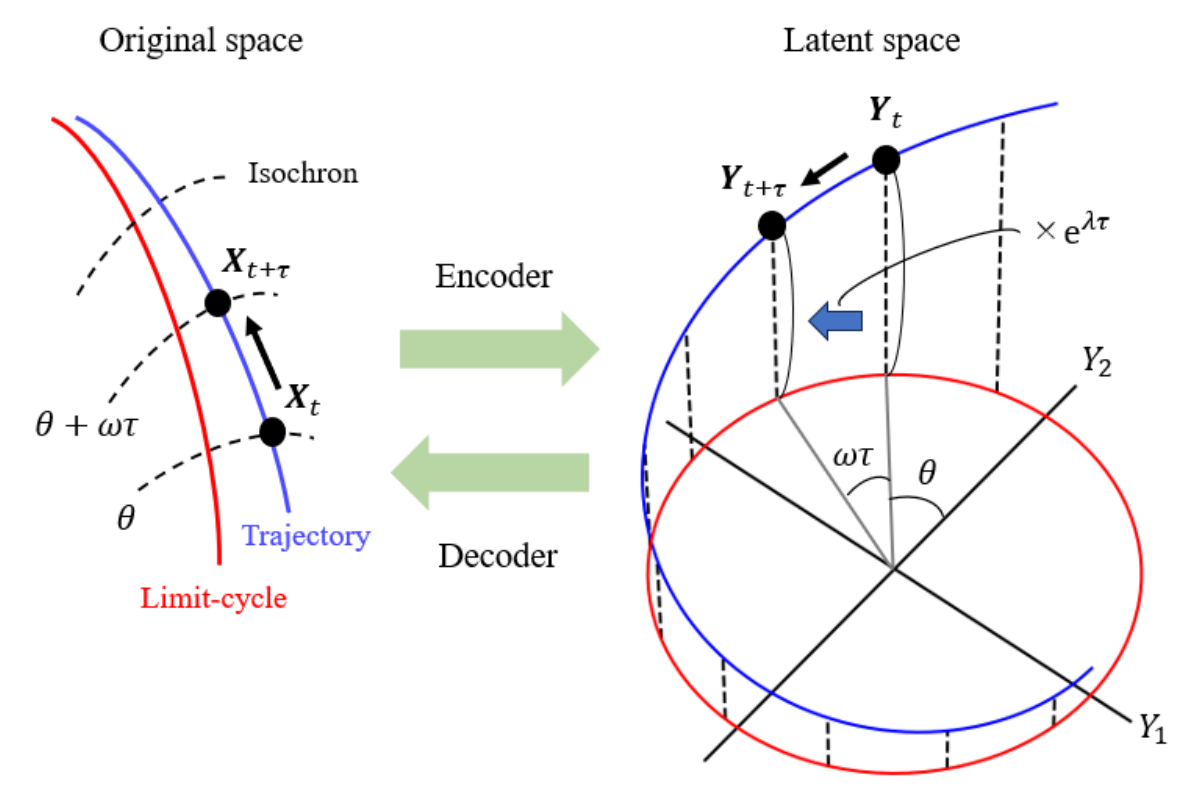}
\caption{
Phase autoencoder. The encoder maps the original state space to a latent space, and the decoder maps the latent space to the original state space. The autoencoder is trained in such a way that a pair of variables in the latent space represent the asymptotic phase of the oscillator, and that the oscillator state on the limit cycle is mapped to a plane in the latent space on which another latent variable is zero. The red line shows the limit cycle, the blue line shows the oscillator orbit, and the dashed line shows the isochron (level set of the asymptotic phase).
}\label{fig:concept}
\end{figure}

The asymptotic phase and PSF are generally not obtainable analytically.
When the differential equation describing the oscillator is known, the PSF can be calculated 
numerically by the adjoint method~\cite{ermentrout1996type}. However, 
since detailed mathematical models are often unavailable for real-world systems,
studies have also been conducted to obtain the asymptotic phase and PSF directly from time-series data
~\cite{ota2009weighted,netoff2012experimentally,imai2017robust,Schmid2010,Kutz2016,Williams2015,lusch2018deep,wilson2016isostable,Namura2022,kralemann2008phase}.

Recently, it has been shown that the asymptotic phase is essentially related to the Koopman eigenfunction of the oscillator with a pure imaginary eigenvalue characterized by the natural frequency~\cite{mauroy2013isostables}, and Dynamic Mode Decomposition (DMD) and its extensions~\cite{Schmid2010,Kutz2016,Williams2015}, which can estimate Koopman eigenvalues and eigenfunctions from time-series data, have been employed to estimate the asymptotic phase in a data-driven manner.
However, there remain challenges with such methods: first, it is not easy to estimate the asymptotic phase for high-dimensional systems, and second, it is difficult to reconstruct the original oscillator state from the phase as an input.

In this study, we propose a machine-learning method for estimating the asymptotic phase based on an autoencoder (Fig.~\ref{fig:concept}). 
The autoencoder is a neural network that encodes original input data into latent variables and then decodes the original input data from the latent variables.
Specifically, we design an autoencoder with a latent space in which a pair of latent variables represent the asymptotic phase of the oscillator, while another latent variable represents the oscillator's deviation from the limit cycle.
This enables the estimation of the asymptotic phase and PSF of the oscillator and reconstruction of the oscillator state in the original space from the phase value in a data-driven manner.
As an application, we propose a simple method for globally synchronizing two oscillators using the trained autoencoder.

This paper is organized as follows. First, we briefly outline the phase reduction method in Sec.~\ref{sec:phase}.
We then describe the proposed phase autoencoder
in Sec.~\ref{sec:method}. 
The validity of the proposed phase autoencoder is illustrated through numerical simulations in Sec.~\ref{sec:experiment},
and a simple method for globally synchronizing two oscillators using the trained autoencoder is presented in Sec.~\ref{sec:synchronization}.
Section~\ref{sec:discussion} discusses the relationship of the latent variables with the Koopman eigenfunctions,
and Sec.~\ref{sec:conclusion} gives conclusions.

\section{PHASE REDUCTION}\label{sec:phase}

\subsection{Limit-cycle oscillator}

We describe the state of a limit-cycle oscillator in terms of a vector ${\bm X} \in \mathbb{R}^{d_X}$, which obeys the following ordinary differential equation (ODE):
\begin{equation}
    \frac{d}{dt}{\bm X}(t) = {\bm F}({\bm X}(t)),
\label{eq:main_ode}
\end{equation}
where $t$ is the time and ${\bm F}({\bm X}) : {\mathbb R}^{d_X} \to  {\mathbb R}^{d_X}$ is a smooth vector field representing the oscillator dynamics. 
We assume that this dynamical system has an exponentially stable limit-cycle solution ${\bm X}_0(t)$ 
of period $T$ in the state space satisfying $ {\bm X}_0(t + T) = {\bm X}_0(t) $, and denote its basin of attraction as $A \subseteq {\mathbb R}^{d_X}$.

\subsection{Asymptotic phase}

To characterize the oscillator state ${\bm X}$, we introduce a phase function $\Theta({\bm X}) : A \to [0, 2\pi]$ that gives the asymptotic phase of ${\bm X}$~\cite{Winfree1980, Kuramoto2003}. 
First, the phase for a state ${\bm X}_0(t)$ on the limit cycle is defined as
\begin{equation}\label{eq:ph1}
    \Theta({\bm X}_0(t)) = (t\bmod T) \omega,
\end{equation}
where $\omega = 2\pi/T$ is the natural frequency.
By this definition, a phase value between $0$ and $2\pi$ is assigned to the state ${\bm X}_0(t)$ on the limit cycle, and this phase value increases constantly with $t$, i.e.,
\begin{equation}\label{eq:phase_fuction}
    \frac{d}{dt}\Theta({\bm X}_0(t))=\omega.
\end{equation}
Note that the point of phase 0 is given by ${\bm X}_0(0)$, which can be chosen arbitrarily.
In what follows, we denote the state at phase $\theta$ on the limit cycle as ${\bm X}_0(\theta)$.
Next, the phase of a state ${\bm X}$ in the basin $A$ is defined to be the same phase value as the state on the limit cycle ${\bm X}_0(\theta)$ 
if they eventually converge to the same state on the limit cycle as shown in Fig.~\ref{fig:concept}.
In this way, the phase function $\Theta$ can be defined so that
\begin{equation}\label{eq:phase_fuction2}
    \frac{d}{dt}\Theta({\bm X}(t))=\omega
\end{equation}
holds for any state ${\bm X}$ in $A$  obeying Eq.~(\ref{eq:main_ode}).

\subsection{Phase sensitivity function}

The gradient of the phase function
\begin{gather}\label{eq:psf}
    {\bm Z}(\theta)=\text{grad}_{{\bm X}={\bm X}_0(\theta)}\Theta({\bm X}) \in {\mathbb R}^{d_X}
\end{gather}
evaluated at ${\bm X}_0(\theta)$ on the limit cycle characterizes the linear response property of the oscillator's phase to small perturbations given at phase $\theta$ and is called the phase sensitivity function (PSF)~\cite{Kuramoto2003} .
It can be seen from Eqs.~(\ref{eq:main_ode}) and (\ref{eq:psf}) that 
\begin{align}
    \frac{d}{dt}\Theta({\bm X}_0(\theta))&=\text{grad}_{{\bm X}={\bm X}_0(\theta)}\Theta({\bm X}))\cdot\frac{d}{dt}{\bm X}_0(\theta)\cr
    &=\text{grad}_{{\bm X}={\bm X}_0(\theta)}\Theta({\bm X}))\cdot {\bm F}({\bm X}_0(\theta))\cr
    &={\bm Z}(\theta)\cdot {\bm F}({\bm X}_0(\theta)).
\end{align}\label{eq:hoge}
Thus, from Eq.~(\ref{eq:phase_fuction}), the following normalization condition for the PSF should be satisfied:
\begin{equation}
    {\bm Z}(\theta)\cdot {\bm F}({\bm X}_0(\theta)) = \omega.
	\label{eq:normalization}
\end{equation}
It is well known that the PSF can be calculated by solving the adjoint equation~\cite{ermentrout1996type}
\begin{align}
\frac{d}{d\theta} {\bm Z}(\theta) = - J(\theta)^{\top} {\bm Z}(\theta),
\end{align}
where $J(\theta) = D{\bm F}({\bm X}_0(\theta))$ is the Jacobian matrix of ${\bm F}({\bm X})$ evaluated at ${\bm X} = {\bm X}_0(\theta)$ and $\top$ denotes matrix transpose.

When the oscillator is subjected to a weak perturbation ${\bm p} \in {\mathbb R}^{d_X}$ and is described by
\begin{align}
\frac{d}{dt} {\bm X}(t) = {\bm F}({\bm X}(t)) + {\bm p}({\bm X}(t), t),
\end{align}
the phase $\theta(t) = \Theta({\bm X}(t))$ of this oscillator state approximately obeys a phase equation
\begin{align}
\frac{d}{dt} \theta(t) = \omega + {\bm Z}(\theta) \cdot {\bm p}({\bm X}(t), t)
\end{align}
at the lowest order in ${\bm p}$.
The simplicity of this phase equation has facilitated detailed analysis of the synchronization dynamics of limit-cycle oscillators~\cite{Winfree1980, Kuramoto2003, nakao2016phase,Ermentrout2010,monga2019phase,kuramoto2019concept,ermentrout2019recent,fukami2024data}.

\section{Phase autoencoder}\label{sec:method}
\begin{figure*}[t]
 \begin{center}
  \includegraphics[scale=0.52]{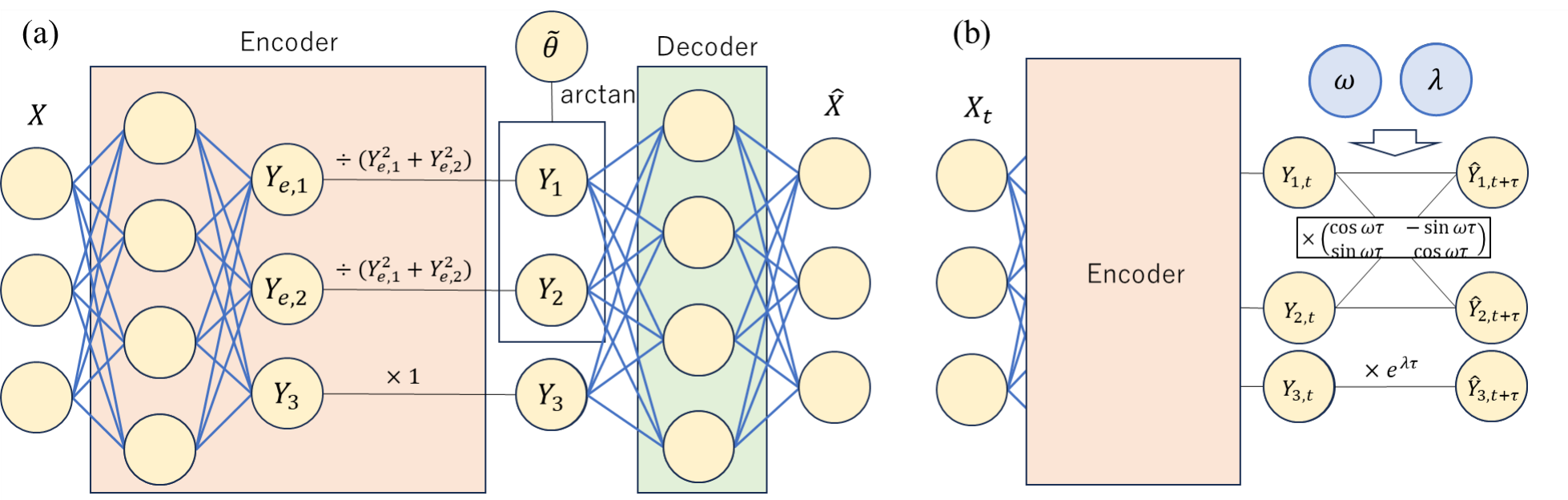}
  \caption{Phase autoencoder architecture. (a) The encoder transforms the inputs into three-dimensional latent variables. The first two variables are normalized and correspond one-to-one with the asymptotic phase. The decoder reconstructs the input from the three latent variables. (b) Time evolution of the three latent variables transformed by the encoder. The blue lines and circles ($\omega$ and $\lambda$) indicate the parameters to be trained.}
  \label{fig:archi}
 \end{center}
\end{figure*}

\subsection{Background}

Before presenting our proposed phase autoencoder, we briefly explain the background of our study.
A closely related idea to our phase autoencoder is the Koopman autoencoder~\cite{lusch2018deep,Mardt2018,Wehmeyer2018,Otto2019,Takeishi2017,Yeung2019,Champion2019,Azencot2020,Li2019,Berman2023,Han2021}.
The Koopman autoencoder trains the autoencoder so that the time evolution of the latent variables is described by a linear mapping representing the evolution of the observables by the Koopman operator.
This is achieved by training the autoencoder to reduce the difference between the latent variables at the next time step predicted by the linear mapping and the latent variables encoded from the system state at the next time step.
In pioneering works, Takeishi {\it et al.}~\cite{Takeishi2017} proposed a method 
for searching the latent space that allows linear mapping, and Lusch {\it et al.}~\cite{lusch2018deep} proposed a method for learning both encoding and linear mapping in the latent space.

Since the asymptotic phase is the argument of the Koopman eigenfunction associated with a pure imaginary eigenvalue (see Sec.~\ref{sec:discussion})~\cite{mauroy2013isostables,shirasaka2017phase}, the Koopman autoencoder could, in principle, be used for evaluating the oscillator's asymptotic phase. 
However, these Koopman autoencoders were not designed to perform phase reduction of limit cycles.
In Ref.~\cite{Namura2022}, a simple method to perform data-driven phase reduction was proposed, which approximates the asymptotic phase based on polynomial regression without relying on the Koopman operator theory.
The drawbacks of the latter method were the difficulties in handling oscillators with higher dimensions ($>2$) and also in reconstructing the oscillator state in the original space from the given phase value.

In this study, we design an autoencoder in such a way that a pair of the latent variables directly represent the asymptotic phase and another latent variable decays with time as the oscillator state approaches the limit cycle.
This allows direct estimation of the asymptotic phase $\theta$ from the oscillator state ${\bm X}$
and reconstruction of the oscillator state ${\bm X}_0(\theta)$ on the limit cycle when the phase $\theta$ is given.
In what follows, we explain the phase autoencoder and its loss functions for learning the above-described latent variables.

\subsection{Proposed Phase Autoencoder}

An autoencoder~\cite{hinton2006reducing,fukami2023grasping} is an artificial neural network 
consisting of an encoder $f_{enc} : {\mathbb R}^{d_X} \to {\mathbb R}^{d_Y}$, which transforms 
an input vector ${\bm X} \in {\mathbb R}^{d_X}$ to
a latent vector ${\bm Y} \in {\mathbb R}^{d_Y}$,
and a decoder $f_{dec} : {\mathbb R}^{d_Y} \to {\mathbb R}^{d_X}$, which approximately reconstructs the original 
vector $\hat{\bm X} \in {\mathbb R}^{d_X}$ from the latent vector ${\bm Y}$. Here, $d_X$ and $d_Y$ are the dimensions of the input space and the latent space, respectively.
In this study, $d_Y$ is fixed at 3 (see discussion in Sec. \ref{sec:discussion}).
The architecture of the autoencoder is shown in the Fig.~\ref{fig:archi} (a).

We assume that the oscillator state ${\bm X}(t)$ is sampled at a sampling time interval of $\tau > 0$, which is denoted as ${\bm X}_t$, and transformed to a latent vector ${\bm Y}_t = f_{enc}({\bm X}_t)$ by the autoencoder. 
We design the autoencoder so that the latent variables (components of the three-dimensional latent vector) ${\bm Y}_t = (Y_{1,t},Y_{2,t},Y_{3,t})$ at time step $t$ satisfy 
\begin{gather}
    Y_{1,t}^2+Y_{2,t}^2=1\label{eq:norm},\\
    Y_{1,t+\tau} = Y_{1,t}\cos(\omega \tau) - Y_{2,t}\sin(\omega \tau)\label{eq:step1},\\
    Y_{2,t+\tau} = Y_{1,t}\sin(\omega \tau) + Y_{2,t}\cos(\omega \tau)\label{eq:step2},\\
    Y_{3,t+\tau} = e^{\lambda\tau}Y_{3,t}, \label{eq:step3}
\end{gather}
where $\omega$ and $\lambda$ are also the parameters to be learned (Fig.~\ref{fig:archi} (b)).
The first three equations for the latent variables $Y_{1,t}$ and $Y_{2,t}$ represent that the oscillator state ${\bm X}_t$ is mapped onto a unit circle on the $Y_{1}-Y_{2}$ plane in the latent space and rotates on it with a constant frequency $\omega$ as the oscillator state evolves. 
Thus, $Y_1$ and $Y_2$ correspond to the asymptotic phase.
The last equation for another latent variable $Y_{3,t}$ represents that the deviation of the oscillator state from the $Y_{1}-Y_{2}$ plane shrinks exponentially with time at a rate $\lambda$.
We stress that we design the autoencoder to satisfy Eqs. (\ref{eq:step1})-(\ref{eq:step3}) for all $\tau>0$.

Since the limit cycle is exponentially stable, we assume that $\lambda$ is negative and the oscillator states on the limit cycle are embedded on the $Y_1 - Y_2$ plane satisfying $Y_3=0$, where we expect that the information around the limit cycle is embedded collectively in $Y_3$.
See Fig.~\ref{fig:concept} for a schematic of our definition of the latent variables.
We will discuss the relationship of this variable $Y_3$ with the amplitude variable~\cite{shirasaka2020phase} defined via the Koopman operator theory in Sec.~\ref{sec:discussion}.

Our phase autoencoder learns the encoding function $f_{enc}$ 
and the decoding function $f_{dec}$ to satisfy Eqs. (\ref{eq:norm})-(\ref{eq:step3}).
To guarantee a one-to-one correspondence between the phase and the two latent variables $Y_1$ and $Y_2$, we internally train an encoder $\tilde{f}_{enc}$ that maps ${\bm X}$ to a non-normalized latent vector ${\bm Y_e}$ as $\bm{Y_e} = (Y_{e,1},\ Y_{e,2},\ Y_{e,3}) = \tilde{f}_{enc}({\bm X})$,  then normalize this ${\bm Y_e}$ as
\begin{gather}
    R = \sqrt{Y_{e,1}^2 + Y_{e,2}^2},\\
    {\bm Y} = (Y_{e,1}/R,\ Y_{e,2}/R,\ Y_3),
\end{gather}
and denote the result after the normalization as the final encoder, ${\bm Y} = f_{enc}({\bm X})$. 
The decoder is represented as
\begin{gather}
    \bm{\hat{X}} = f_{dec}({\bm Y}),
\end{gather}
which takes the normalized latent vector ${\bm Y}$ as the input and output the oscillator state $\bm{\hat{X}}$ approximating the original state ${\bm X}$.

In addition to the autoencoder, we also introduce a neural network $f_{step}$ that learns the dynamics of the latent vector so that
\begin{equation}
    {\bm Y}_{t+\tau} = f_{step}({\bm Y}_{t})
\end{equation}
holds approximately. 
Specifically, $f_{step}$ learns the parameters $\omega$ and $\lambda$ from the discretized dynamics in the latent space described by Eqs.~\eqref{eq:step1}-\eqref{eq:step3}.

Using the trained encoder, the phase function can be written as
\begin{equation}
    \Theta({\bm X}) = \arctan \left( \frac{Y_2 }{ Y_1 } \right) = \arctan \left( \frac{ f_{enc}({\bm X})_{2} }{ f_{enc}({\bm X})_{1} } \right),
\end{equation}
where the subscript $i=1,2$ indicates the $i$th component of the vector-valued function $f_{enc}$.
This indeed gives the asymptotic phase of the oscillator, because it satisfies $\Theta({\bm X}_{t+\tau}) = \Theta({\bm X}_t) + \omega \tau$ from Eqs.~\eqref{eq:step1} and \eqref{eq:step2}, which gives $\dot\Theta({\bm X}_t) = \omega$ in the $\tau \to 0$ limit,
provided that the autoencoder is successfully trained.
We can also estimate the PSF by
\begin{align}
    {\bm Z}(\theta) &= \text{grad}_{{\bm X}}\Theta(\bm{X})|_{{\bm X}=f_{dec}(\theta)},
\end{align}
where the gradient can be easily computed by using automatic differentiation of the machine learning framework~\cite{paszke2017automatic}.

\subsection{Loss functions}

To train the autoencoder, we introduce three types of loss functions: reconstruction loss, dynamics loss (phase and deviation), and auxiliary loss.
The reconstruction loss of the autoencoder is the distance between the input data and the reconstructed output,
\begin{gather}
    L_{recon} = {\mathbb E} \left[ \left\|{\bm X}_t - f_{dec}(f_{enc}({\bm X}_t)) \right\|^2 \right],
    \label{eq:reconloss}
\end{gather}
where $\| \cdots \|$ represents the Euclidean norm and ${\mathbb E}[ g({\bm X}_t) ] = (1/T_{max}) \sum_{t=0}^{T_{max}-1} g({\bm X}_t)$ represents the average of a quantity $g({\bm X}_t)$ over the input data $\{ {\bm X}_t | t=1, ..., T_{max} \}$ of length $T_{max}$.
Here, $T_{max}$ is the maximum number of inputs that can be obtained from a single orbit, determined by the simulation length and its computation step width.

The dynamics loss is introduced to control the property of the latent space.
We consider $k$-step evolution ($k=1, \ldots, K$) of the latent variables by the neural network $f_{step}$ and introduce the following loss functions characterizing the prediction errors for the latent variables corresponding to the phase and deviation: 
\begin{gather}
    L_{pha} = 
    {\mathbb E}\left[
    \sum_{k=1}^{K} \alpha_{k}
    \sum_{i=1}^{2} \left|f_{enc}({\bm X}_{t+k\Delta t})_i-f_{step}^k(f_{enc}({\bm X}_t))_i \right|^2
    \right],
    \\
    L_{dev} = 
    {\mathbb E}\left[
    \sum_{k=1}^{K} \alpha_{k}
    \sum_{i=3}^{d_Y} \left| f_{enc}({\bm X}_{t+k\Delta t})_i-f_{step}^k(f_{enc}({\bm X}_t))_i \right|^2
    \right],
\end{gather}
where the latent variables are summed componentwise (the subscript $i$ indicates the $i$th component), and $\Delta t$ is the sampling interval of the input data.
This $\Delta t$ needs to be scaled appropriately to take account of the period $T$ of the limit cycle.
In training, $\Delta t$ is fixed, but it is expected that Eqs. (\ref{eq:step1})-(\ref{eq:step3}) are satisfied for any $\tau>0$ if trained with sufficient data.

In the above equations, the coefficient $\alpha_k$ represents the weight of the prediction error for each time step $k$ and is set as
\begin{equation}
    \alpha_k = 1/k^{\text{min}(1.0, L_{pha})}.
\end{equation}
Note that $\alpha_k$ is updated at each epoch of learning (`epoch' is a single learning of the entire training dataset).
The reason for choosing the weight $\alpha_k$ as above is as follows. In the early stage of the learning where small-step predictions of the phase are not accurate ($L_{pha} > 1$), the above $\alpha_k$ emphasizes the prediction errors for small $k$ to accelerate the learning. As the learning progresses and $L_{pha}$ becomes smaller, $\alpha_k$ approaches $1$ and prediction errors for larger $k$ are also taken into account.

Since the loss function $L_{pha}$ has a local solution that represents a false stationary state in the latent space corresponding to $\omega=0$, we also introduce the following loss function to avoid such an inappropriate solution:
\begin{align}
    L_{aux}
     &=
    {\mathbb E} \left[
      \left( \frac{1}{B} \sum_{b=1}^B f_{enc}({\bm X}^b_t)_1 \right)^2 + \left( \frac{1}{B} \sum_{b=1}^B f_{enc}({\bm X}^b_t)_2 \right)^2
	\right],
\end{align}
which characterizes the deviation of the center of mass of the latent variables $(Y^b_{1},\ Y^{b}_2) = (f_{enc}({\bm X}^b)_1,\ f_{enc}({\bm X}^b)_2 )$ on the $Y_{1} - Y_{2}$ plane in the latent space from the origin.
Here, $B$ is the batch size of the learning and ${\bm X}^b$ and $Y^{b}_i$ denote the input data and latent variables in a batch, respectively (`batch' is a small subset of the entire dataset used in the learning).
Since each $(Y^b_{1},\ Y^{b}_2)$ exists on a unit circle in the latent space by the normalization, $L_{aux}$ takes smaller values when these variables distribute evenly on the unit circle within a batch.
We use this $L_{aux}$ in the early stage of learning to avoid local solutions and turn it off as the learning progresses.

Using the above three types of loss functions, we introduce the following total loss function:
\begin{equation}
    L = w_{recon} L_{recon} + w_{pha} L_{pha} + w_{amp} L_{dev} + w_{aux} L_{aux},
\end{equation}
where $w_{recon}$, $w_{pha}$, $w_{amp}$, and $w_{aux}$ are weight parameters controlling relative contributions from the individual loss functions.
We minimize this total loss function to train our phase autoencoder.

\section{Numerical Experiments}\label{sec:experiment}

\subsection{Training datasets and parameters}

We evaluated the proposed phase autoencoder using four types of limit-cycle oscillators as examples. We considered the Stuart-Landau oscillator, for which the analytical solution of the phase function is known, the FitzHugh-Nagumo model, a fast-slow oscillator with a more complex phase function, the Hodgkin-Huxley model, a $4$-dimensional oscillator realistically describing spiking neurons, and a ring network of excitable FitzHugh-Nagumo units exhibiting a traveling pulse, an example of a high-dimensional oscillator.

The data for all of these oscillators were generated by direct numerical simulations of the models.
The datasets for the training were generated as follows.
First, we evaluated the limit-cycle solution and the period $T$. We then generated the training data and determined the the training parameters.
We note that the parameters of the training datasets were the same for all oscillators.
We calculated the standard deviation of each component of the state vector of the oscillator over one period on the limit cycle as
\begin{gather}
    \sigma_i = \sqrt{\frac{1}{N_s}\sum_{j=0}^{N_s-1} \left( {\bm X}_0(\frac{2\pi j}{N_s})_i\right)^2-\left(\frac{1}{N_s}\sum_{j=0}^{N_s-1}
     {\bm X}_0(\frac{2\pi j}{N_s})_i
     \right)^2}
\end{gather}
for $i=1, ..., d_X$, where $N_s$ is the number of sample points on the limit cycle,
and randomly chose $N_s$ initial points ${\bm X}_{s,j}$ as
\begin{gather}
    {\bm X}_{s,j} = {\bm X}_0(\frac{2\pi j}{N_s}) + \gamma_2\bm{\sigma}\odot \bm{\xi}~(j=0,1,\cdots,N_s-1),
    \cr
    \bm{\xi}\sim N_{d_X}(0, 1),
\end{gather}
where ${\bm \sigma} = (\sigma_1, ..., \sigma_{dX})$, $\odot$ denotes the Hadamard (element-wise) product, and $N_{d_X}(0,1)$ denotes the $d_X$-dimensional normal distribution.
We then evolved these initial points for $\gamma_1T$.

In the above setting, $N_s$, $\gamma_1$, and $\gamma_2$ are the parameters for data generation, which affect the estimation accuracy of the autoencoder.
For example, when $\gamma_2=0$, the autoencoder could not accurately estimate the asymptotic phase near the limit cycle and, consequently, it was difficult to estimate the phase sensitivity function.
Also, when $\gamma_1$ was too large, most of the data were on the limit cycle, which adversely affected the learning and did not capture the relaxation dynamics.
In all of the following numerical experiments, unless otherwise noted, we used the same set of parameter values, 
$N_s = 1000$, $\gamma_1 = 3$, and $\gamma_2 = 0.5$, respectively.

Our autoencoder has a standard structure~\cite{hinton2006reducing} except for the 
normalization of the latent variables $Y_1$ and $Y_2$, which consists of linear transformations, ReLU~\cite{nair2010rectified}, and batch normalization~\cite{ioffe2015batch} (Fig.~\ref{fig:archi}).
The encoder has two hidden layers, each with 100 units, and the decoder has three hidden layers, each with 100 units.
The epoch size was $50$ and the learning rate was $0.001$.
The weights of individual loss functions were $w_{recon}=1.0, w_{pha}=0.5, w_{amp}=0.5$, and $w_{aux}=2.0$, and the maximal step of evolution by the neural network $f_{step}$ was $K=20$.
Additionally, we varied $w_{pha}$ and $w_{aux}$ to $w_{pha}=5.0$ and $w_{aux}=0.0$ after the initial stage of training, 
i.e., when the loss functions satisfied $L_{pha}<0.01$ and $L_{aux}<0.05$.
The inputs were normalized by the standard deviations of the individual variables as a preprocessing of the data from the limit-cycle oscillators.
We used the standard machine-learning library PyTorch~\cite{paszke2019pytorch} for the actual implementation of our phase autoencoder. 

\subsection{Stuart-Landau oscillator}

First, we applied our method to the Stuart-Landau (SL) oscillator~\cite{Kuramoto2003,stuart1960non,landau1944comptes},
which is a normal form of the supercritical Hopf bifurcation described by
\begin{gather}
\frac{d}{dt}\begin{bmatrix}
x_1\\
x_2\\
\end{bmatrix}=
\begin{bmatrix}
x_1 - \alpha x_2 - (x_1-\beta x_2)(x_1^2+x_2^2)\\
\alpha x_1 + x_2 - (\beta x_1+x_2)(x_1^2+x_2^2)\\
\end{bmatrix},
\end{gather}
where $x_1$ and $x_2$ are the variables and $\alpha$ and $\beta$ are the parameters.
The limit-cycle solution is $(x_1(t), x_2(t)) = (\cos \omega t, \sin \omega t)$ if we choose $(1,0)$ as the initial condition, which is a unit circle on the $x_1 - x_2$ plane, where the natural frequency is $\omega = \alpha - \beta $. We set $\alpha=2\pi$ and $\beta=1$ in the numerical experiment, which gives the oscillation period $T \simeq 1.19$.
We trained the autoencoder and estimated the phase function of the SL oscillator.

We set the number of initial points as $n = 1000$, the number of maximal evolution steps as $K = 20$, and the training sample interval as $\Delta t = 0.05$. 
Here, we chose $K$ and $\Delta t$ so that the product of $K$ and $\Delta t$ roughly coincides with the oscillation period $T$.
We also used the same values for $n$ and $K$ in the subsequent numerical experiments.

\begin{figure}[htbp]
\includegraphics[scale=0.24]{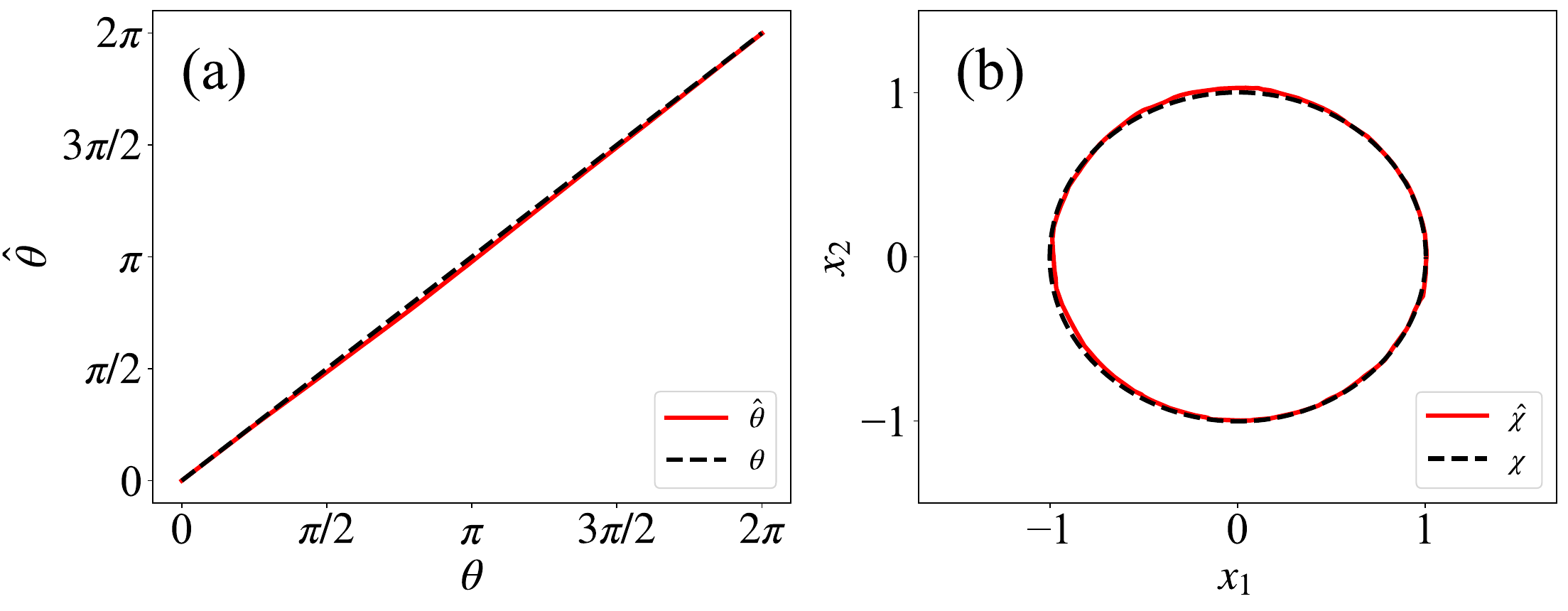}
\caption{
(a) Comparison of the estimated phase (red solid line) $\hat{\theta}$ and the true phase (black dotted line) $\theta$ of the SL oscillator. 
(b) Reconstructed limit cycle $\hat{\chi}$ by the autoencoder (red solid curve) compared with the true limit cycle $\chi$ (black dotted curve).
}
\label{fig:SL_LC}
\end{figure}

First, we examined whether the correct phase values were assigned to the oscillator states on the limit cycle.
We sampled the oscillator state ${\bm X}_0(\theta)$ on the limit cycle with regular intervals of the phase $\theta$.
We then estimated the phase values of the sampled oscillator states by the autoencoder to confirm if the phase function is properly embedded.
Figure~\ref{fig:SL_LC} (a) compares the original phase $\theta$ and the estimated phase $\hat{\theta}$ from ${\bm X}_0(\theta)$, showing that the asymptotic phase is accurately estimated by the trained autoencoder.
We also verified that the limit cycle is embedded in the latent space on a plane with $Y_3 = 0$.
Figure~\ref{fig:SL_LC} (b) shows the decoder's output of the limit cycle $\hat{\chi}$ calculated by setting $Y_3=0$ and sampling $(Y_1, Y_2)$ at regular intervals of the phase $\theta$ as $\bm{Y}=(\cos \theta, \sin \theta,0)$ and compare it with the original limit cycle (unit circle) $\chi$ of the SL oscillator, showing good agreement.

\begin{figure}[htbp]
\includegraphics[scale=0.20]{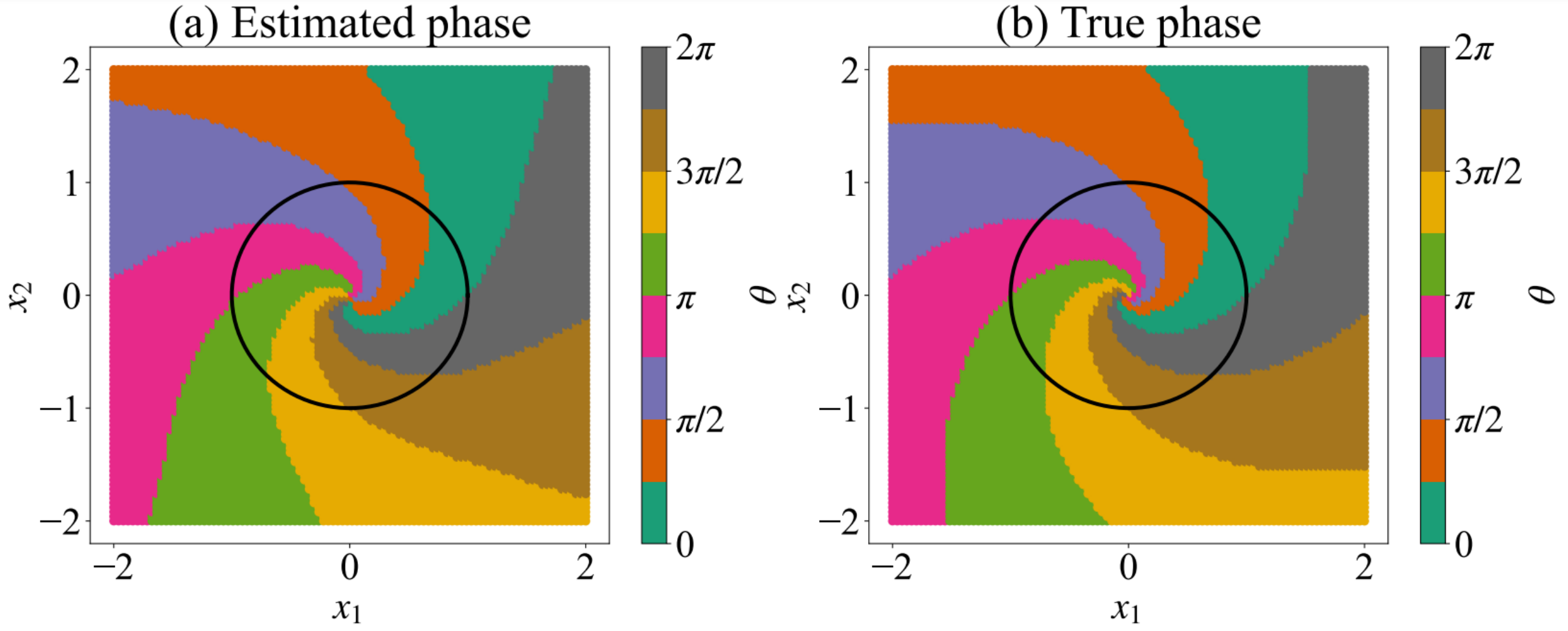}
\caption{Asymptotic phase of the Stuart-Landau oscillator. (a) Estimated phase function by the autoencoder; (b) True phase function.
The colors represent the phase value from $0$ to $2\pi$ (discretized for visual clarity), where $(x_1,x_2) = (1,0)$ is chosen as the origin of the phase with $\theta=0$.
The black circle in each figure represents the limit cycle.}
\label{fig:SL_PF}
\end{figure}

Next, we examined the asymptotic phase estimated by the autoencoder.
The true asymptotic phase function of the SL oscillator is analytically given by~\cite{nakao2016phase}
\begin{gather}
    \Theta(x_1, x_2) = \text{arctan}\left(x_2/x_1 \right) - \log \sqrt{x_1^2+x_2^2}.
\end{gather}
Figure~\ref{fig:SL_PF} compares the estimated phase function with the true phase function
on the $x_1 - x_2$ plane.
It can be seen that the trained autoencoder reproduces the phase function well around the limit cycle. 
For the oscillator states far from the limit cycle, the estimation accuracy is degraded due to lack of training data in these regions.

\begin{figure}[htbp]
\includegraphics[scale=0.24]{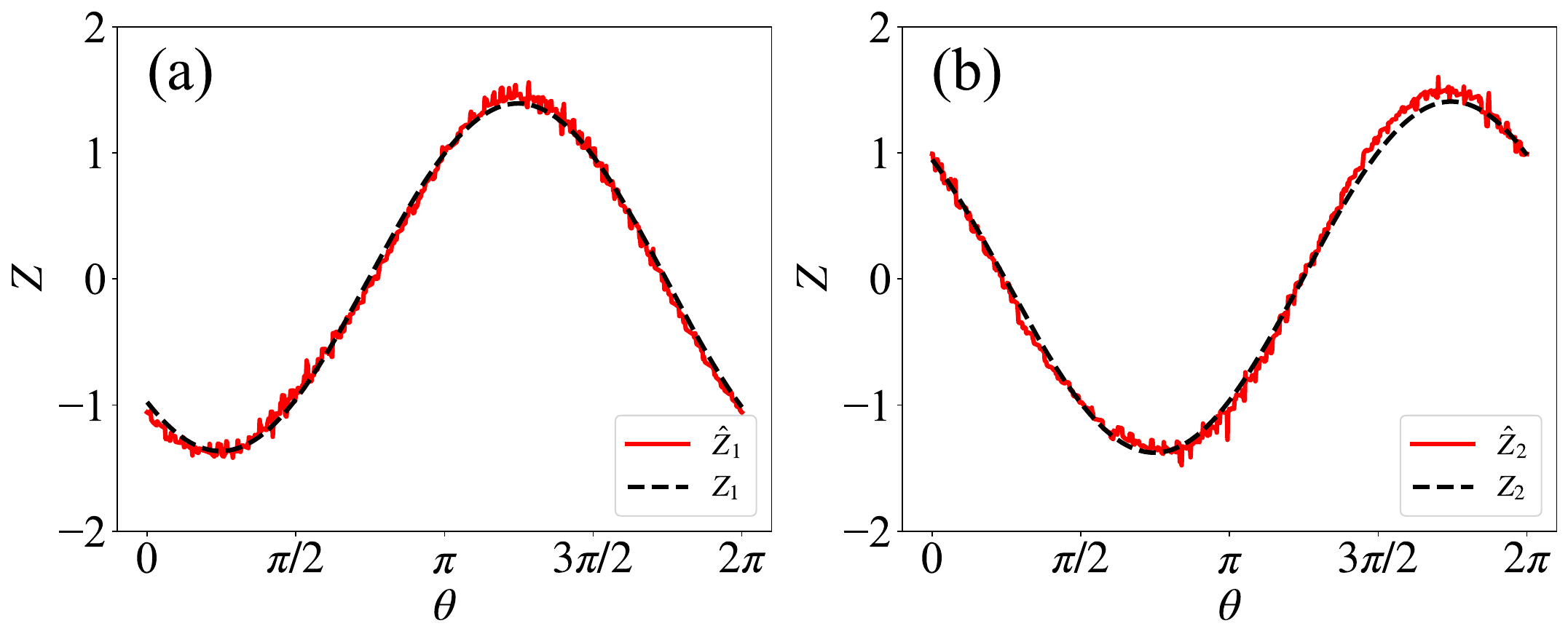}
\caption{Phase sensitivity function (PSF) of the SL oscillator. Each figure shows the estimated PSF (red solid curve) and the true PSF (black dashed curve). (a) $x_1$ component; (b) $x_2$ component. 
}
\label{fig:SL_PSF} 
\end{figure}

We also compared the estimated PSF with the true PSF in Fig.~\ref{fig:SL_PSF}, showing good agreement.
Here, the true PSF is analytically obtained from the true asymptotic phase as~\cite{nakao2016phase}
\begin{align}
	{\bm Z}(\theta) 
	&= ( Z_1(\theta), Z_2(\theta) )
	\cr
	& = ( - \sin \theta - \beta \cos \theta,\ \cos \theta - \beta \sin \theta).
\end{align}
Since the PSF is a derivative of the phase function, the results were sensitive to small errors in the estimated phase function and therefore exhibited small fluctuations around the true curves.

\begin{figure}[htbp]
\includegraphics[scale=0.22]{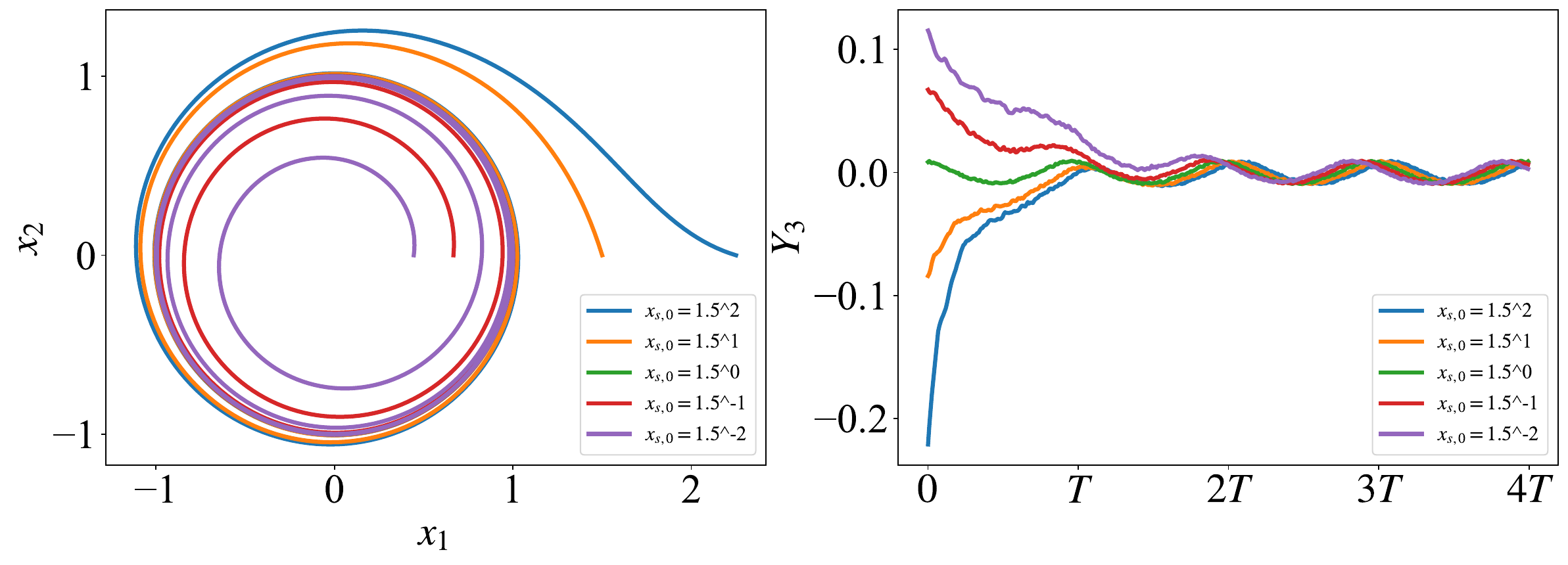}
\caption{
Time evolution of the latent variable $Y_3$ for five trajectories with different initial points of the SL oscillator. (a) Dynamics in the original state space. (b) Time evolution of the latent variable $Y_3$.}
\label{fig:SL_R}
\end{figure}

Finally, we describe the evolution of the third latent variable $Y_3$ characterizing the deviation of the oscillator state from the limit cycle.
Figure~\ref{fig:SL_R} shows how $Y_3$ evolved for five different initial states converging to the limit cycle.
We observe that $Y_3$ decayed to zero roughly exponentially as intended. However, it did not converge to zero accurately but exhibited small-amplitude oscillations due to estimation errors of the trained autoencoder.

\subsection{FitzHugh-Nagumo model}

As the second example, we applied the proposed method to the FitzHugh-Nagumo (FHN) model~\cite{Winfree1980,Ermentrout2010} described by
\begin{gather}
\frac{d}{dt}\begin{bmatrix}
x_1\\
x_2\\
\end{bmatrix}=
\begin{bmatrix}
x_1-x_1^3/3-x_2+I\\
\epsilon(x_1+a-bx_2)\\
\end{bmatrix},
\end{gather}
where $x_1,x_2$ are the variables and $\epsilon=0.08,a=0.7, b=0.8$, and $I=0.8$ are the parameters.
This system has a stable limit cycle of period $T \simeq 36.56$ with fast-slow dynamics.
We set the number of initial points as $n = 1000$, the  number of maximal evolution steps as $K = 20$, and the training sample interval as $\Delta t = 1.8$.

\begin{figure}[htbp]
\includegraphics[scale=0.24]{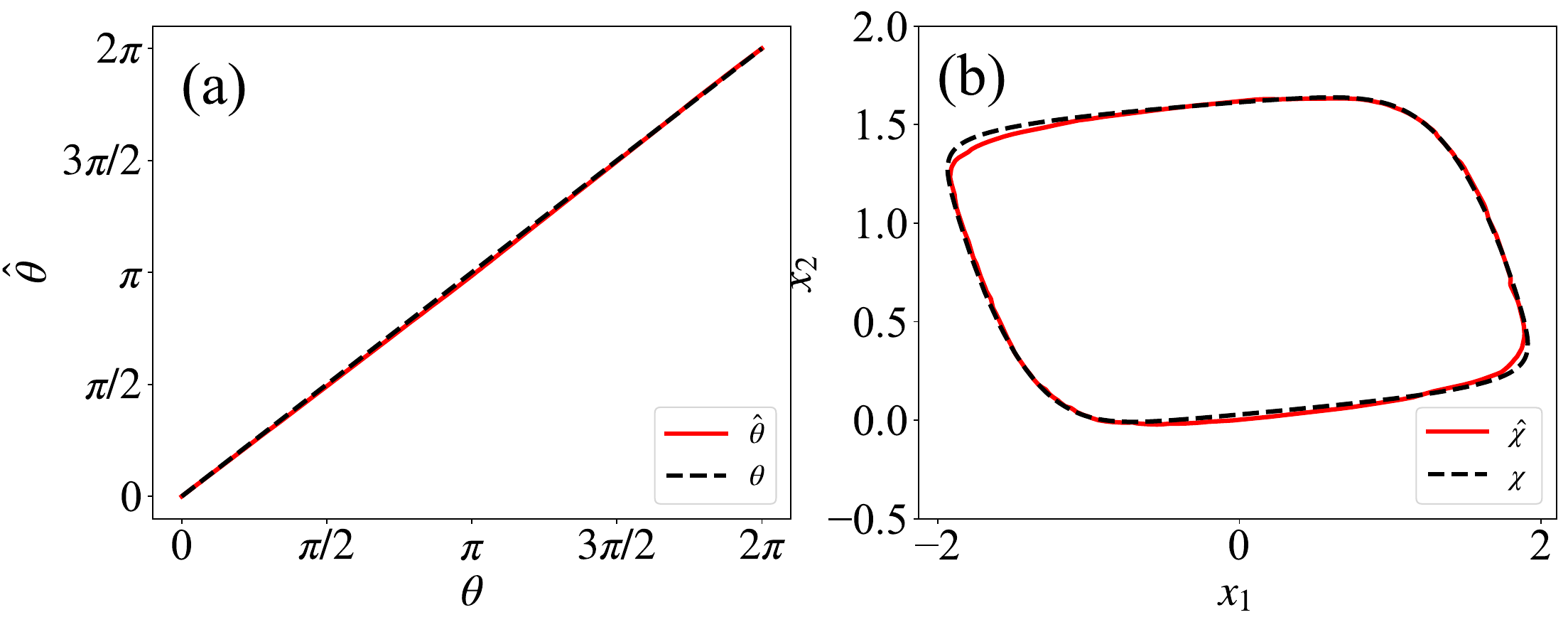}
\caption{(a) Comparison of the estimated phase $\hat{\theta}$ (red solid line) and the true phase $\theta$ (black dotted curve) of the FHN model. 
(b) Reconstructed limit cycle $\hat{\chi}$ by the autoencoder (red solid curve) compared with the true limit cycle $\chi$ (black dotted curve).}\label{fig:FHN3_LC} 
\end{figure}

As in the case of the SL oscillator, we first examined if the phase values were properly assigned to the states on the limit cycle and if the limit cycle was embedded in the plane with 
$Y_3=0$ in the latent space.
As Fig.~\ref{fig:FHN3_LC} shows, the phase on the limit cycle was accurately learned by the autoencoder. 
The limit cycle was also reconstructed reasonably well, but it slightly shifted inward from the original limit cycle near the nullclines.
We presume that the cause of this small shift is that the training data generated by our method were slightly biased inside the true limit cycle. 

\begin{figure}[htbp]
\includegraphics[scale=0.18]{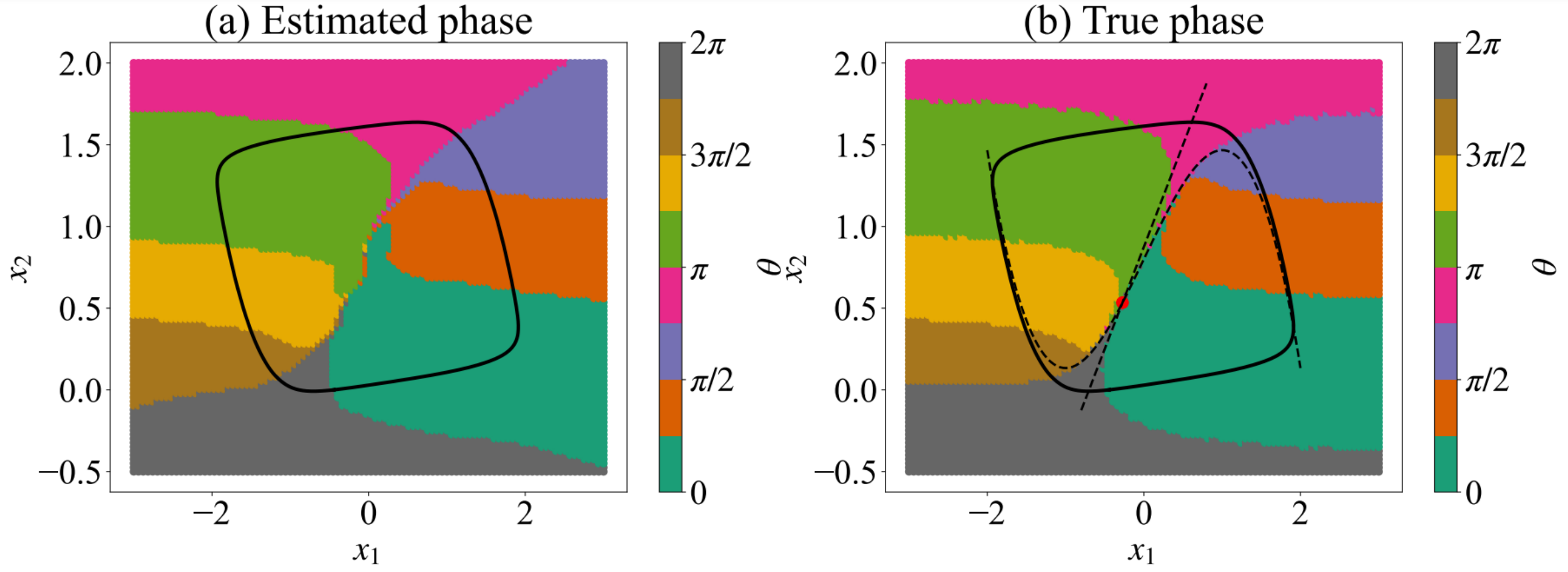}
\caption{Asymptotic phase of the FHN model. (a) Estimated phase by the autoencoder. (b) Analytical phase value.  The colors represent the phase value from $0$ to $2\pi$ (discretized for visual clarity), where $(x_1,x_2) = (-0.445, 0)$ is chosen as the origin of the phase ($\theta=0$).
The black curve in each figure represents the limit cycle.
}
\label{fig:FHN3_PF}
\end{figure}

Next, we estimated the asymptotic phase using the autoencoder and compared with the result obtained by direct numerical calculation
in Fig.~\ref{fig:FHN3_PF}.
The autoencoder reproduced the phase function near the limit cycle well, but there were discrepancies around the unstable fixed point inside the limit cycle and also in the regions away from the limit cycle.
This is because the training data did not contain the trajectories passing through these regions and therefore the autoencoder could not learn the dynamics in these regions. 
In practice, because the limit cycle is strongly attracting in general, the oscillator state is almost always near the limit cycle and rarely visits these regions. 
Thus, these discrepancies are not a serious problem in using the trained autoencoder for analyzing and controlling synchronization.

\begin{figure}[htbp]
\includegraphics[scale=0.20]{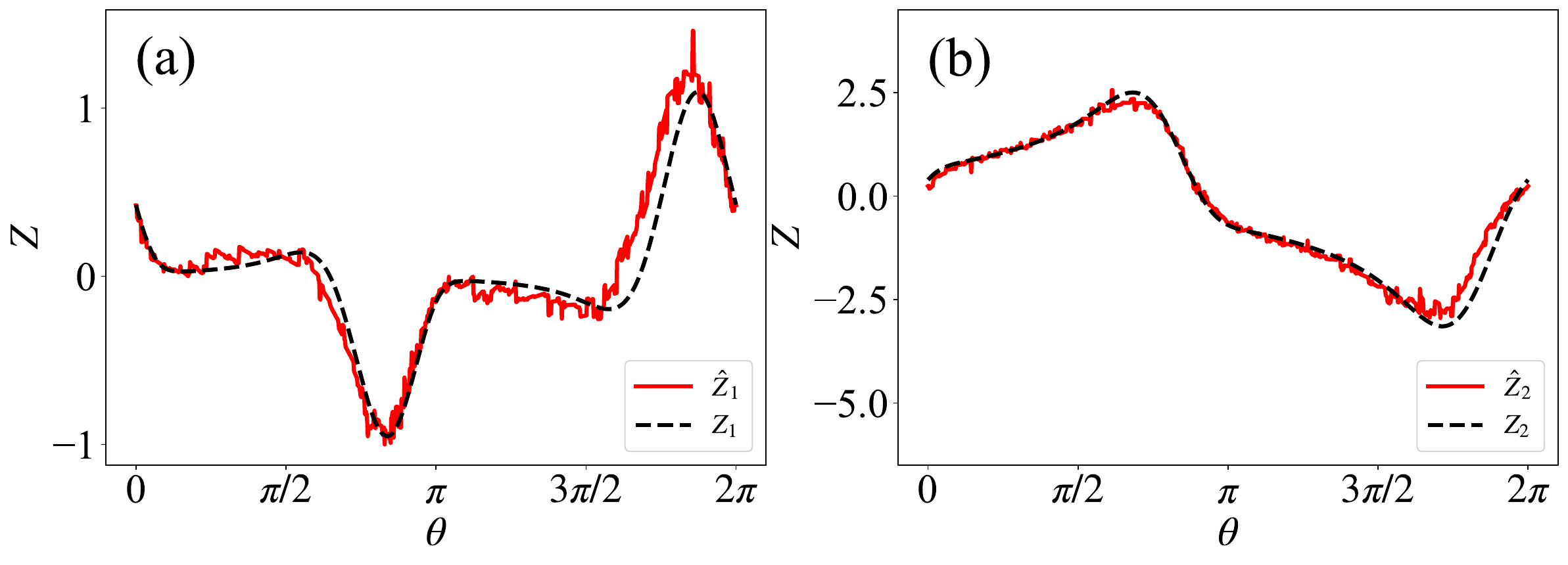}
\caption{Phase sensitivity function (PSF) of the FHN model. Each figure shows the estimated FHN (red solid curve) and the true PSF (black dashed curve). (a) $x_1$ component; (b) $x_2$ component. True PSF was calculated by solving the adjoint equation.}\label{fig:FHN3_PSF}
\end{figure}

Finally, the PSF ${\bm Z} = (Z_1, Z_2)$ estimated by the autoencoder and the true PSF obtained from the adjoint equation are compared in Fig.~\ref{fig:FHN3_PSF}. Since estimating the derivatives enhances noise, the estimated PSF was jagged, but still 
mostly reproduced the true PSF; reflecting the discrepancy of the trajectories near the nullclines, we also observe discrepancies in the PSF near the nullclines.


\subsection{Hodgkin-Huxley model}

\begin{figure}[htbp]
\includegraphics[scale=0.19]{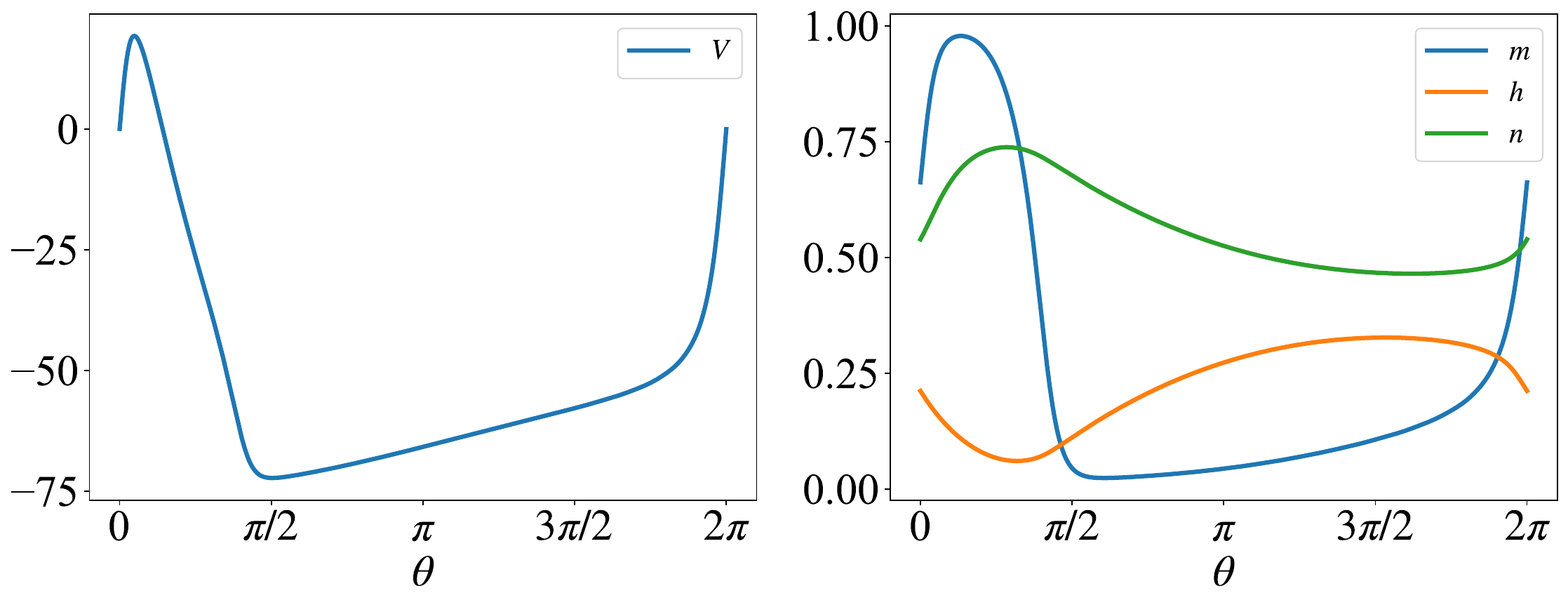}
\caption{
Evolution of the state variables on the limit cycle of the Hodgkin-Huxley model. Membrane potential $V$ (left) and other channel variables $m$, $h$, and $n$ (right) vs. phase. 
}\label{fig:HH_X}
\end{figure}

As the third example, we applied the proposed method to the Hodgkin Huxley (HH) model of spiking neurons~\cite{hodgkin1952quantitative}. The HH model realistically describes how the action potentials in neurons are generated and is given by
\begin{gather}
\frac{d}{dt}\begin{bmatrix}
V\\
m\\
h\\
n\\
\end{bmatrix}=
\begin{bmatrix}
(G_{Na}m^3h(E_{Na}-V) + G_Kn^4(E_K-V) \\+ G_L(E_L-V) + I)/C\\
\alpha_m(V)(1.0-m)-\beta_m(V)m\\
\alpha_h(V)(1.0-h)-\beta_h(V)h\\
\alpha_n(V)(1.0-n)-\beta_n(V)n\\
\end{bmatrix},
\end{gather}
where $V$ is the membrane potential, $m, h, n$ are the channel variables, and the voltage-dependent rate constants are given by
\begin{align}
\alpha_m(V) &= 0.1 (V+40)(1-\text{exp}(-(V+40)/10)), \cr
\beta_m(V) &= 0.1 \text{exp}(-(V+65)/18), \cr
\alpha_h(V) &= 0.07 \text{exp}(-(V+65)/20), \cr
\beta_h(V) &= 1/(1+\text{exp}(-V+35)/10)), \cr
\alpha_n(V) &= 0.01 (V+55)/(1-\text{exp}(-(V+55)/10)), \cr
\beta_n(V) &= 0.125 \text{exp}(-(V+65)/80).
\end{align}
We set the parameters as $C=1.0, G_{N_a}=120.0, G_K=36.0, G_L=0.3, E_{N_a}=50.0, E_K=-77.0$, and $E_L=-54.4$,
for which the HH model possessed a stable limit cycle of period  $T \simeq 10.12$.
For the training, we set the number of initial points as $n = 1000$, the number of maximal evolution steps as $K = 20$, and the sampling interval as $\Delta t = 0.5$.
Figure~\ref{fig:HH_X} shows the evolution of the variables $V$, $m$, $h$, and $n$ for one oscillation period.

\begin{figure}[htbp]
\includegraphics[scale=0.19]{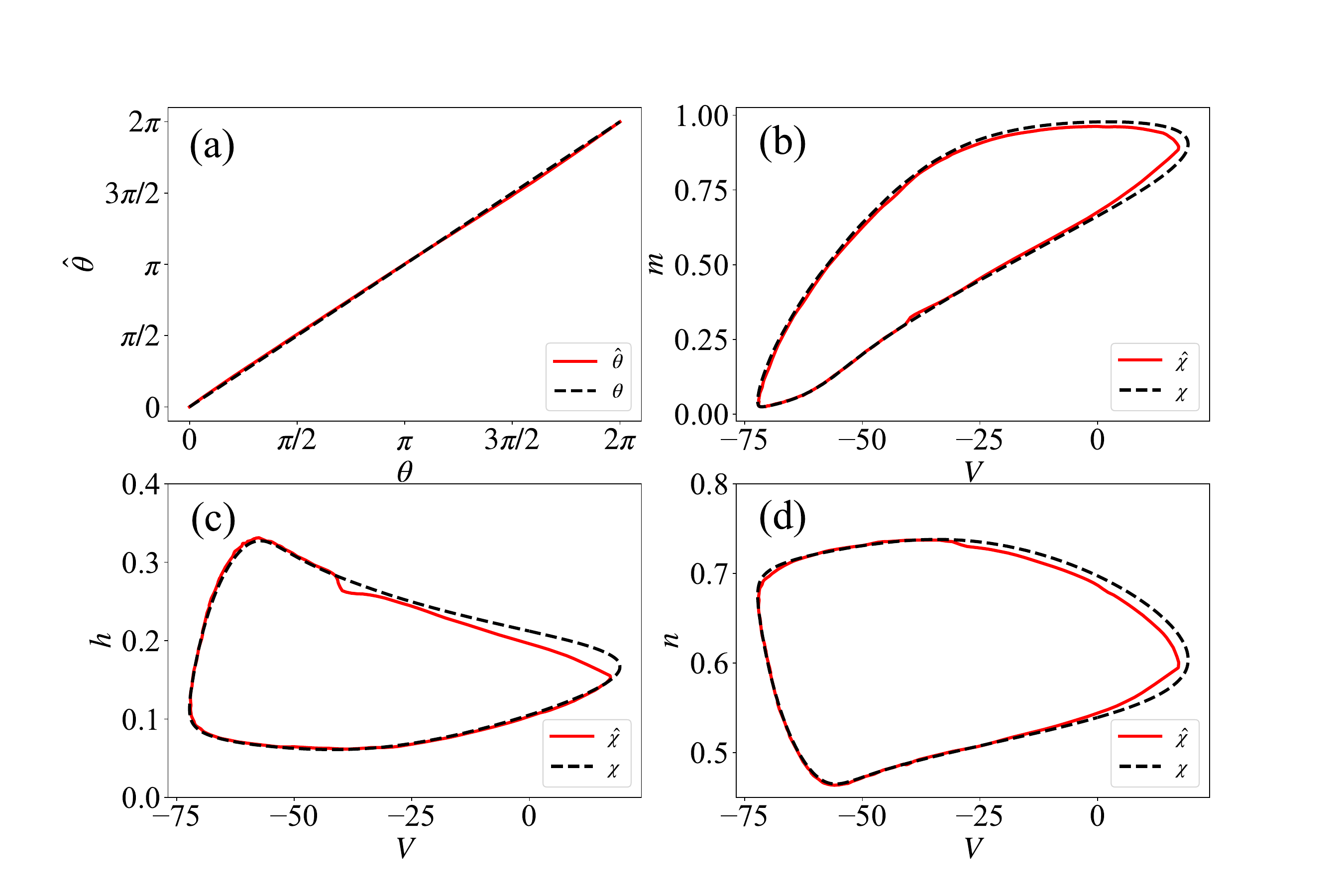}
\caption{(a) Comparison of the estimated phase $\hat{\theta}$ (red solid line) and the true phase $\theta$ (black dotted curve) of the HH model. 
(b-d) Reconstructed limit cycle $\hat{\chi}$ projected on (b) $V - m$, (c) $V - h$, and (d) $V - n$ planes (red solid curves) compared with the true limit cycle $\chi$ (black dotted curves). }
\label{fig:HH_LC}
\end{figure}

As in the previous cases, we confirmed that appropriate phase values were assigned to the states on the limit cycle (Fig.~\ref{fig:HH_LC} (a)).
We also examined if the limit cycle could be reconstructed from the embedded representation.
Since the HH model is $4$-dimensional, we projected the trajectory on the plane spanned by
$V$ and one of other variables, $m$ (Fig.~\ref{fig:HH_LC} (b)), $h$ (Fig.~\ref{fig:HH_LC} (c)), or $n$ (Fig.~\ref{fig:HH_LC} (d)).
The reconstructed limit cycle agreed with the original one reasonably well, but, as in the case with the FHN model,  it was somehow shifted inward from the original limit cycle.

\begin{figure}[htbp]
\includegraphics[scale=0.19]{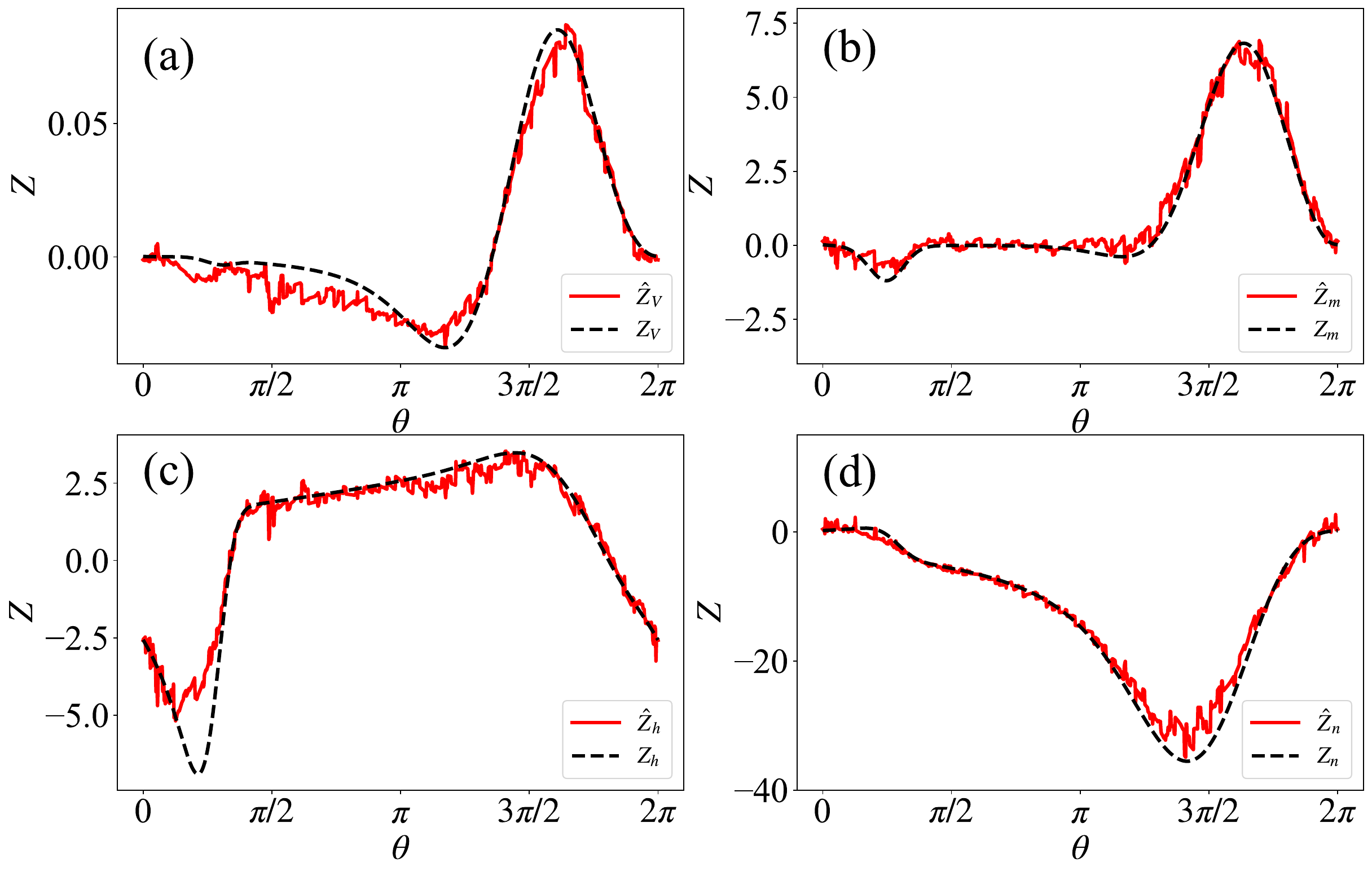}
\caption{Phase sensitivity function (PSF) of the HH model. Each figure shows the estimated PSF (red solid curve) and the true PSF (black dashed curve). (a) $V$ component; (b) $m$ component; (c) $h$ component; (d) $n$ component. The true PSF was calculated by solving the adjoint equation.}\label{fig:HH_PSF}
\end{figure}

The PSF ${\bm Z} = (Z_V, Z_m, Z_h, Z_n)$ estimated by the autoencoder is compared with the true PSF obtained from the adjoint equation in Fig.~\ref{fig:HH_PSF}, showing reasonably good agreement; the deviations in the $h$ component from the true value near $\theta=\pi/4$ is due to the difficulty in estimating the asymptotic phase around this point because of the sudden change in $V$.
We note that, in physiological experiments, only the $V$ component of the HH model can be stimulated and thus the $V$ component of the PSF is important practically; the PSF with respect to the channel variables $m$, $h$, and $n$ were calculated here to verify the validity of our trained autoencoder.

\subsection{Collectively Oscillating Network}

As the final example, we applied the proposed method to a collectively oscillating network (CON) model as an example of a higher-dimensional limit cycle~\cite{nakao2018phase,mircheski2023phase}.
This model consists of multiple FitzHugh-Nagumo elements as the nodes coupled over the links, described by
\begin{gather}
\frac{d}{dt}\begin{bmatrix}
u_i\\
v_i\\
\end{bmatrix}=
\begin{bmatrix}
\epsilon(x_1+a-bx_2)\\
x_1-x_1^3/3-x_2+I\\
\end{bmatrix}
+\sum_{j=1}^NA_{ij}
\begin{bmatrix}
0\\
v_j - v_i\\
\end{bmatrix}
\end{gather}
for $(i=1,2,\cdots,N)$, where $N$ is the number of elements, $u_i, v_i$ are the variables of each element, and $\epsilon=0.08, a=0.7, b=0.8$, and $I=0.32$ are the parameters.
We consider a ring network with $N=10$ and assume that the adjacency matrix is given by
\begin{gather}
A_{ij} = 
\begin{cases}
+0.3 & \text{if }i=j+1, \\
-0.3 & \text{if }i=j-1,\\
0 & \text{(otherwise),}
\end{cases}
\end{gather}
for $i, j = 1, ..., 10$, where $i=11$ is identified with $i=1$ and $i=0$ is identified with $i=10$. This system has a stable limit cycle of period $T \simeq 17.7$. 

\begin{figure}[htbp]
\includegraphics[scale=0.22]{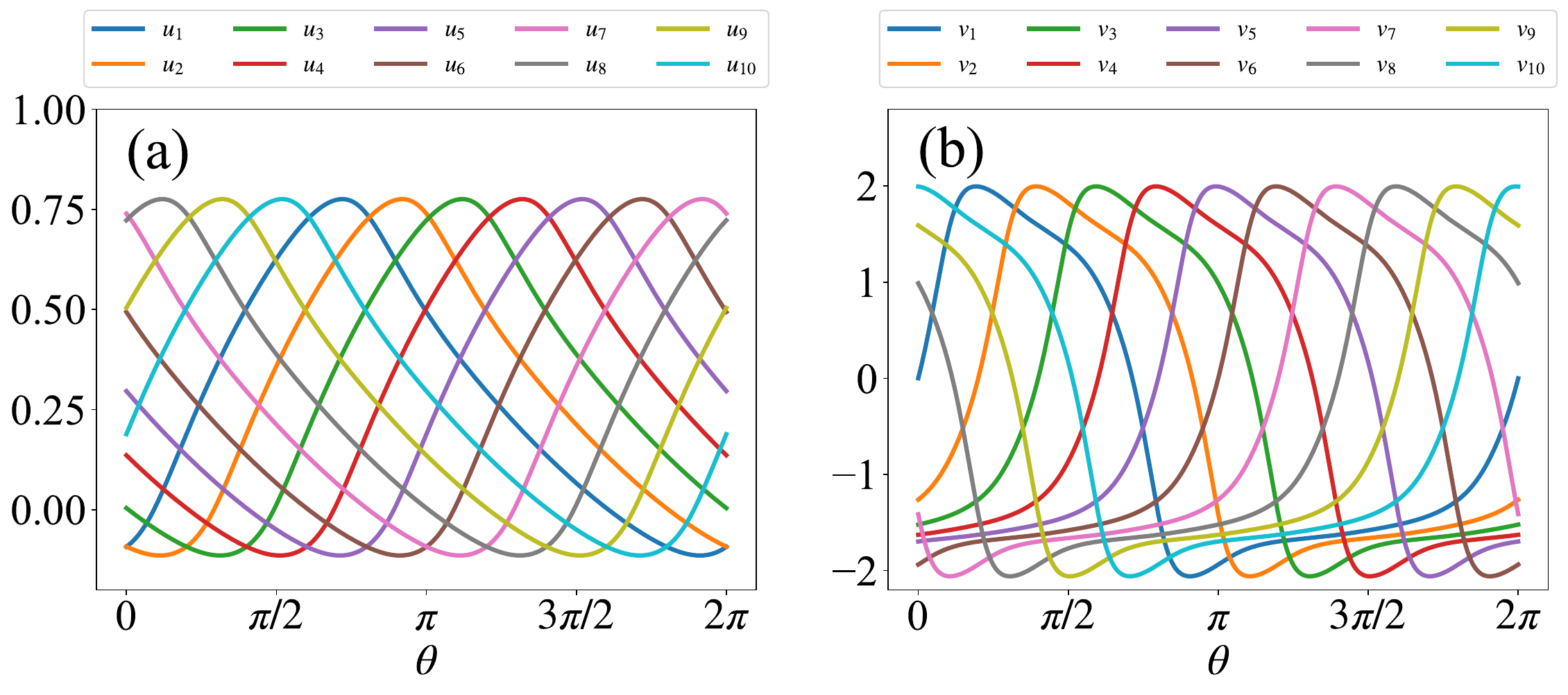}
\caption{Limit cycle of the CON model. Evolution of the state variables for one oscillation period. (a) $u_i$ and (b) $v_i$ $(i = 1, 2, \cdots, 10)$ vs. phase.}
\label{fig:FHNR_X}
\end{figure}

For the training, we set the number of initial points as $n = 1000$, the number of maximal evolution steps as $K = 20$, and the sampling interval as $\Delta t = 0.888$.
Figure~\ref{fig:FHNR_X} shows the evolution of the variables $u_i$ and $v_i(i=1,2,\cdots,10)$ for one oscillation period. 
The stable limit cycle of this model corresponds to a traveling pulse rotating around the ring network~\cite{mircheski2023phase}.

\begin{figure}[htbp]
\includegraphics[scale=0.20]{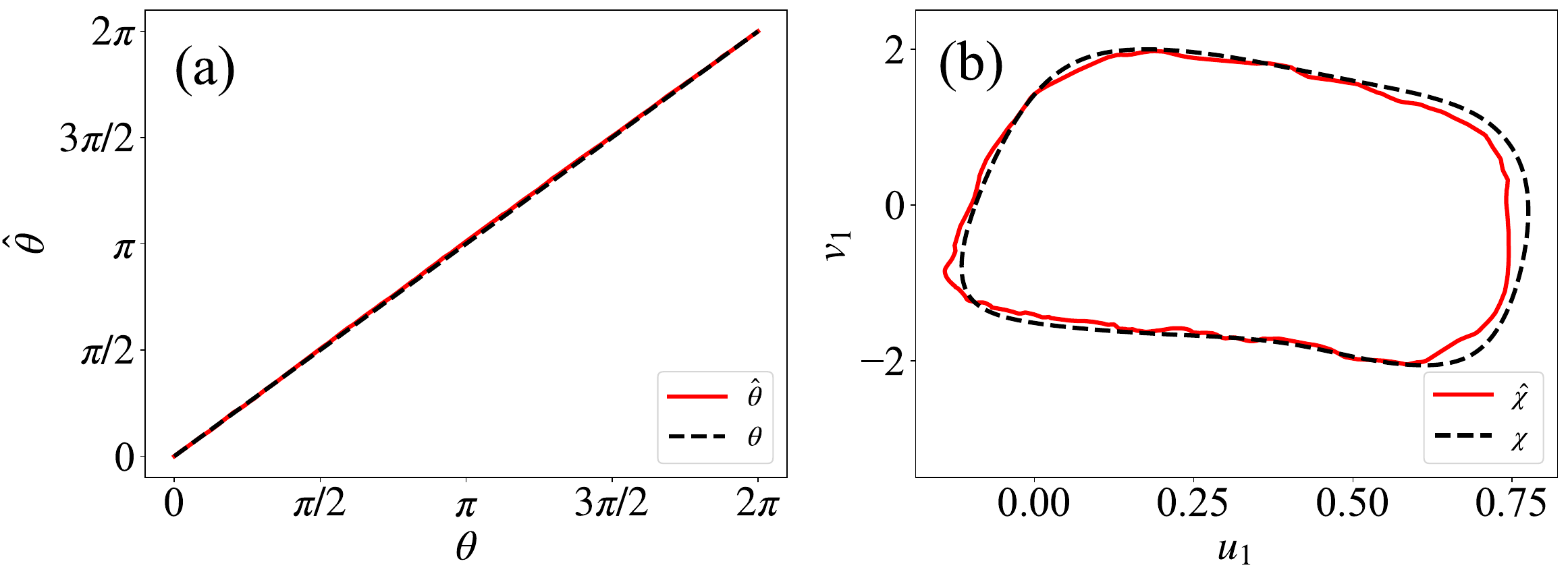}
\caption{(a) Comparison of the estimated phase $\hat{\theta}$ (red solid line) and the true phase $\theta$ (black dotted curve) of the CON model. (b) Reconstructed limit cycle $\hat{\chi}$ projected on $u_1 - v_1$ plane (red solid curve) compared with the true limit cycle $\chi$ (black dotted curve).}\label{fig:FHNR_LC}
\end{figure}

As in the previous cases, we verified that the state on the limit cycle was assigned suitable phase values (Fig.~\ref{fig:FHNR_LC} (a)). We also examined if the limit cycle could be reconstructed from the embedded representation. Since the CON model is 20-dimensional, we showed only the trajectory on the plane spanned by $u_1$ and $v_1$ (Fig.~\ref{fig:FHNR_LC} (b)).
The reconstructed limit cycle agreed with the original one reasonably well, but, as in the case with FHN model and HH model, the state embedded at $Y_3 = 0$ was somehow shifted inward from the original limit cycle.

\begin{figure}[htbp]
\includegraphics[width=\hsize]{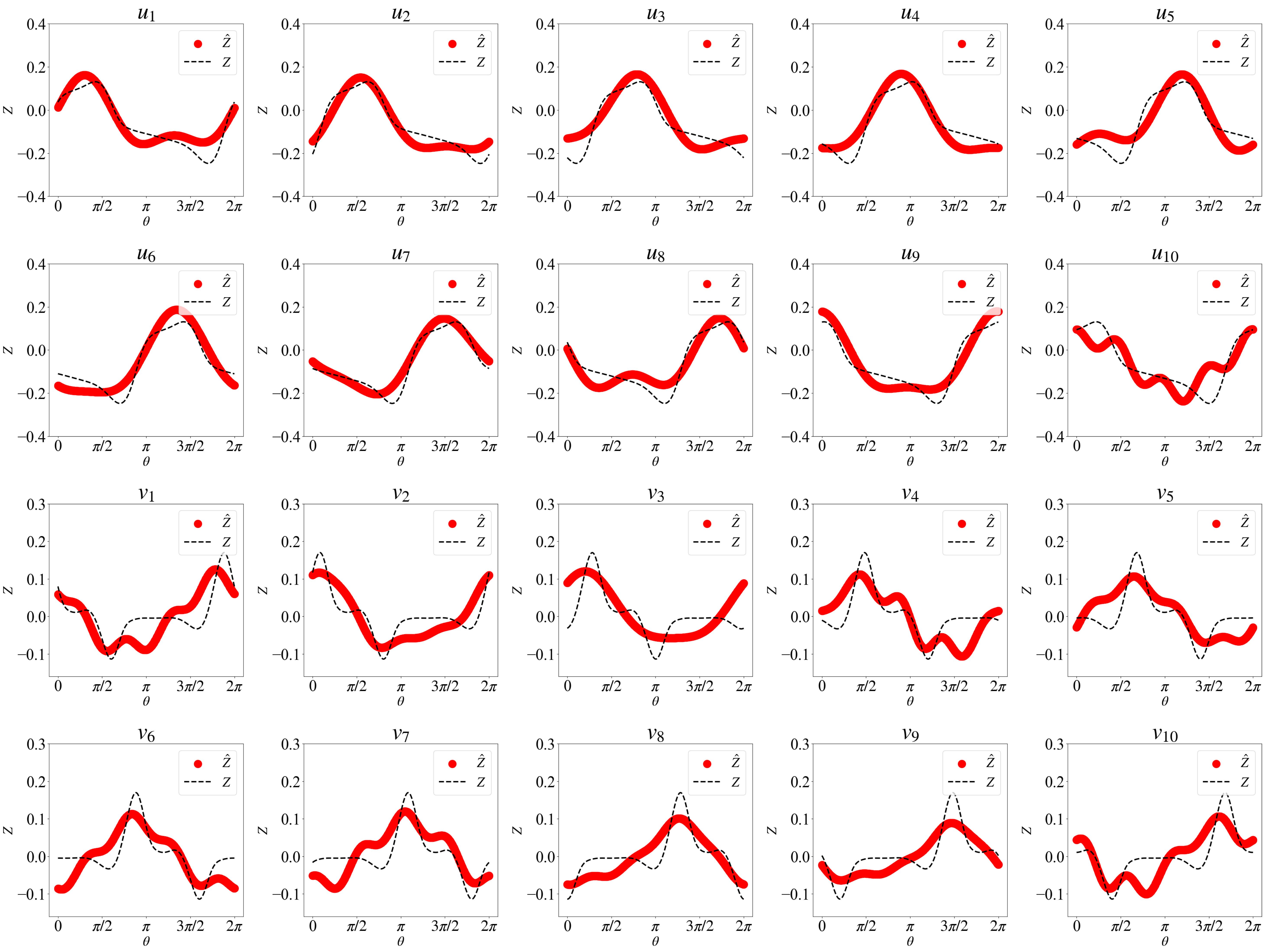}
\caption{Phase sensitivity function (PSF) of the CON model. Each figure shows a single component ($u_1, ..., u_{10}$ or 
  $v_1, ..., v_{10}$) of the estimated PSF (red points) and the true PSF (black dashed curve).
}
\label{fig:FHNR_PSF}
\end{figure}

The PSF ${\bm Z} = (Z_{u_1}, ..., Z_{u_{10}}, Z_{v_1}, ..., Z_{v_{10}})$ with respect to the variables $u_1, ..., u_{10}$ and $v_1, ..., v_{10}$ estimated by the autoencoder is compared with the true PSF obtained from the adjoint equation in Fig.~\ref{fig:FHNR_PSF}.
For the PSF estimated by the autoencoder, the plotted data points are smoothed by the Fourier approximation truncated up to the $5$th modes to reduce fluctuations.
Each component of the PSF is roughly estimated, though considerable deviations from the true curve are observed. This is because the system's dimension is high, $d_X = 20$, and the autoencoder requires a large amount of data for accurate training. 
In practice, however, we typically need only the first several Fourier modes of the PSF in the phase-reduction analysis, and the results obtained here could already be used for explaining synchronization dynamics of the CON model.

\section{Autoencoder-aided synchronization}\label{sec:synchronization}

Here, as an application of the trained phase autoencoder, we present a simple method for globally synchronizing two oscillators.
This method gives a coupling function between the two oscillators that yields a simple pair of sinusoidally coupled phase oscillators when the phase reduction is performed, which has a single in-phase synchronized state as a stable fixed point.

We consider two weakly coupled oscillators with identical properties described by
\begin{align}
    \frac{d}{dt}{\bm X}_1 &= {\bm F}({\bm X}_1) + \epsilon \bm{G}_{12}({\bm X}_1, {\bm X}_2),
    \cr
    \frac{d}{dt}{\bm X}_2 &= {\bm F}({\bm X}_2) + \epsilon \bm{G}_{21}({\bm X}_2, {\bm X}_1), \label{eq:sync}
\end{align}
where $\epsilon$ is a coupling intensity and ${\bm G}_{ij} : {\mathbb R}^{d_X} \times {\mathbb R}^{d_X} \to {\mathbb R}^{d_X}$ is a coupling function.
Using the phase autoencoder, we couple these two oscillators by the following coupling function:
\begin{gather}
    \theta_i = f_{enc}({\bm X}_i),
    \\
    \bm{G}_{ij}({\bm X}_i, {\bm X}_j) = \sin(\theta_j - \theta_i) \frac{\partial}{\partial \theta_i} f_{dec}(\theta_i),
\end{gather}
where $(i, j) = (1, 2)$ or $(2, 1)$.

The reduced phase equation is then given by
\begin{align}
    \frac{d}{dt} \theta_i
    &= {\bm Z}(\theta_i) \cdot\frac{d}{dt}{\bm X}_i\cr
    &= {\bm Z}(\theta_i) \cdot({\bm F}({\bm X}_i) + \epsilon
    \bm{G}_{ij}({\bm X}_i, {\bm X}_j))\cr
    &= \omega + \epsilon {\bm Z}(\theta_i)\cdot
    \bm{G}_{ij}({\bm X}_i, {\bm X}_j)\cr
    &= \omega + \epsilon 
    \sin(\theta_j-\theta_i) \left({\bm Z}(\theta_i) \cdot \frac{\partial}{\partial \theta_i} f_{dec}(\theta_i) \right)\cr
    &= \omega + \epsilon 
    \sin(\theta_j-\theta_i),
    \label{eq:Kuramoto}
\end{align}
where we used 
${\bm Z}(\theta_i) \cdot \frac{\partial}{\partial \theta_i} f_{dec}(\theta_i)=1$
from the normalization condition, Eq.~\eqref{eq:normalization}, 
because ${\bm X}_0(\theta_i) = f_{dec}(\theta_i)$ if the decoder can accurately
reconstruct the oscillator state on the limit cycle.
Since the phase difference $\phi = \theta_1 - \theta_2$ obeys $\frac{d}{dt} \phi = - 2 \epsilon \sin \phi$, the two oscillators are expected to converge to the mutually in-phase synchronized state $\phi=0$ as $t \to \infty$ from any initial phase difference $\phi(0)$ except for $\pi$.

\begin{figure}[htbp]
\includegraphics[scale=0.2]{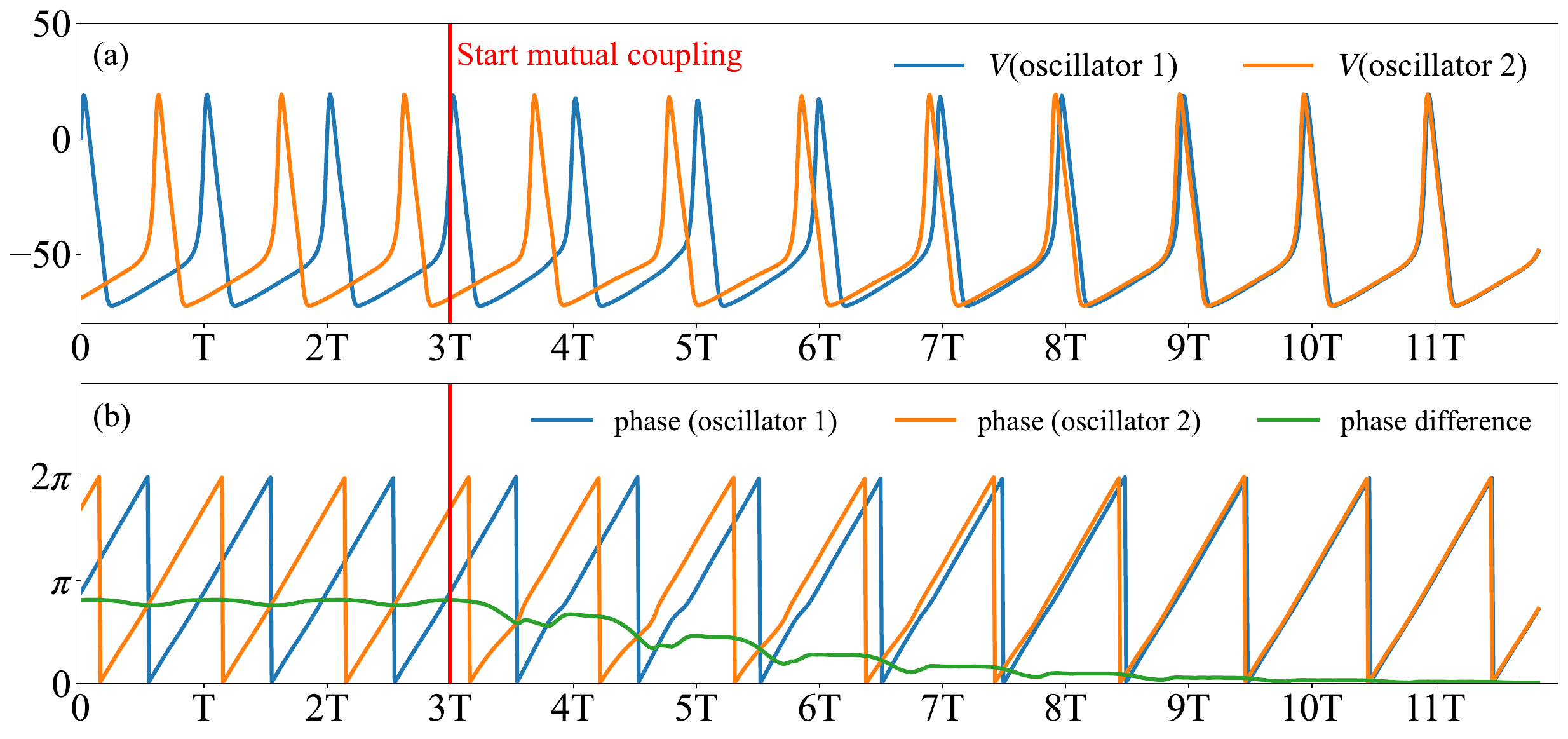}
\caption{Autoencoder-aided mutual synchronization of two HH models. (a) Evolution of the membrane potential $V$ of the HH models 1 and 2. The red line shows the time after which the mutual coupling is turned on. (b) Estimated phases and their difference. The blue line shows the phase of the oscillator 1, the orange line shows the phase of the oscillator 2, and the green line shows the phase difference.}\label{fig:HH_SYNC}
\end{figure}

We here illustrate the autoencoder-aided phase synchronization using the HH model.
We prepared two HH models exhibiting limit-cycle oscillations with the same parameters as before, and introduced the mutual coupling from $t=3T$ with the intensity $\epsilon=0.05$ as in  Eq.~\eqref{eq:sync}. 
Figure~\ref{fig:HH_SYNC} (a) shows the time evolution of the HH models, and (b) shows the estimated phases and their phase difference.
We can confirm that the two HH models are synchronized with each other when the mutual coupling is introduced, confirming the validity of our proposed coupling scheme.

\section{Discussion}\label{sec:discussion}

In our phase autoencoder, we assumed that the latent variables $Y_1$ and $Y_2$ represent the asymptotic phase and rotate on a unit circle with a constant frequency $\omega$, and that another latent variable $Y_3$ represents the overall deviation of the oscillator state from the limit cycle and exponentially decays to zero with a constant rate $\lambda$. Here, we discuss their relationship with the Koopman eigenfunctions of the oscillator.

The Koopman operator~\cite{mauroy2013isostables,mauroy2016global,shirasaka2017phase} $K^{\tau}$ ($\tau \geq 0$) of Eq.~\eqref{eq:main_ode} is defined as $K^{\tau} g({\bm X}(t)) = g({\bm X}(t+\tau))$ for general observables $g : {\mathbb R}^{d_X} \to {\mathbb C}$.
It describes the evolution of observables in the function space and is a linear operator even if the underlying dynamics of ${\bm X}$ is nonlinear. For limit-cycle oscillators, the phase-amplitude reduction frameworks have been developed recently based on the Koopman operator theory~\cite{wilson2016isostable,mauroy2016global,mauroy2018global,shirasaka2017phase};
it is well known that the asymptotic phase and amplitude can be defined by using the principal Koopman eigenfunction associated with the Floquet exponent of the limit cycle.

First, the latent variables $Y_{1}$ and $Y_{2}$ approximately represent the Koopman eigenfunction with a pure imaginary exponent $i \omega$ (eigenvalue $e^{i \omega \tau}$), because if we define $\psi({\bm X}) = Y_{1} + i Y_{2}$, it satisfies the eigenvalue equation $K^{\tau} \psi({\bm X}) = e^{i \omega \tau} \psi({\bm X})$ from Eqs.~\eqref{eq:step1}-\eqref{eq:step3}. 
Next, for two-dimensional oscillators with $d_X = 2$, the latent variable $Y_3$ approximately corresponds to the Koopman eigenfunction with a negative exponent $\lambda$ (eigenvalue $e^{\lambda \tau}$), that is, if we define $r({\bm X}) = Y_3$, it satisfies $K^{\tau} r({\bm X}) = e^{\lambda \tau} r({\bm X})$ and characterizes the decay of the deviation (amplitude) of the oscillator state from the limit cycle.
Though we did not consider in the present study, for higher-dimensional oscillators, we could assume the dimension $d_Y$ of the latent space to be larger than $3$ and similarly introduce latent variables $Y_i$ $ (i=4,\cdots,d_Y)$ that exponentially decay with the rates $\lambda_i$ as $Y_{i,t+\tau} = e^{\lambda_i \tau}Y_{i, t}$. They will then correspond to the other Koopman eigenfunctions with exponents $\lambda_i$ (eigenvalues $e^{\lambda_i \tau}$), provided that $\lambda_i$ are real. If they are complex, $Y_i$ will not correspond to the Koopman eigenfunctions directly, but they still characterize the deviation of the oscillator state from the limit cycle.

In this study, we focused solely on the asymptotic phase represented by $Y_1$ and $Y_2$, and introduced only a single variable $Y_3$ for characterizing the overall deviation from the limit cycle even for the oscillators with $d_X > 2$. 
Also, though we assumed that $Y_3$ decays to $0$ exponentially in Eq.~\eqref{eq:step3}, the trained autoencoder did not accurately reproduce it as can be seen in Fig.~{\ref{fig:SL_R}} because of the limited training data.
Therefore, the latent variable $Y_3$ in our phase autoencoder does not directly correspond to the Koopman eigenfunction with the slowest decaying exponent used in the phase-amplitude reduction theory~\cite{shirasaka2017phase}.
However, it still captures the deviation of the oscillator state from the limit cycle as can be observed in Figs.~\ref{fig:SL_LC}(b),~\ref{fig:FHN3_PF}(b),~\ref{fig:HH_PSF}, and~\ref{fig:FHNR_LC}  showing that the limit cycles are reasonably reconstructed by setting $Y_3=0$. 
%
Thus, though Eq.~\eqref{eq:step3} is not strictly satisfied by the trained phase autoencoder, which is actually not easy to realize in practice, it suffices our purpose of learning the asymptotic phase from time-series data.

Our phase encoder can be regarded as a kind of physics-informed machine learning, which takes advantage of
incorporating physical properties into the learning models~\cite{raissi2019physics,karniadakis2021physics}.
In particular, there are many studies whose target is Hamiltonian systems~\cite{greydanus2019hamiltonian,chen2022learning,toth2019hamiltonian}.
Recently, Toth {\it et al.}~\cite{toth2019hamiltonian} proposed a variational autoencoder whose latent variables correspond to generalized coordinate and momentum,
and Daigavane {\it et al.}~\cite{daigavane2022learning} proposed a network whose latent variables correspond to the action and angle variables of Hamiltonian systems.
Considering the formal resemblance of the phase-amplitude variables of dissipative limit-cycle oscillators with a decaying amplitude and constantly increasing phase to the action-angle coordinates of Hamiltonian oscillators with a constant action variable and constantly increasing phase, we may interpret the proposed phase autoencoder in this study as a kind of physics-informed network similar to Hamiltonian neural networks.

\section{Conclusions}\label{sec:conclusion}

We presented a phase autoencoder whose latent variables represent the asymptotic phase of a limit-cycle oscillator.
We have shown that the asymptotic phases and the phase sensitivity functions can be accurately estimated by the trained autoencoder for several types of low-dimensional limit-cycle oscillators. 
For high-dimensional network, the phase values were assigned on the limit cycle reasonably well.
However, the phase sensitivity function could not be estimated accurately, though its characteristics were qualitatively captured.
This is due to insufficient data, accuracy of learning, and also limitation of the learning model,
and further improving the accuracy of the phase autoencoder for high-dimensional systems is a future problem.
As an application, we proposed a global synchronization method that uses the trained autoencoder
demonstrated its effectiveness by a numerical simulation of the Hodgkin-Huxley model.

\begin{acknowledgments}
We thank N.~Namura for useful discussion.
H.N. acknowledges financial support from JSPS KAKENHI (Nos. JP22K11919 and JP22H00516) and JST CREST (No. JP-MJCR1913).
K.T. acknowledges financial support from the US AFOSR (grant number FA9550-21-1-0178) and the Vannevar Bush Faculty Fellowship (grant number N00014-22-1-2798).
\end{acknowledgments}

\section*{Data Availability Statement}
The data that support the findings of this study are available within the article.
The source codes used for generating the figures will be made available to the public with the publication of this paper.

\bibliography{phaseautoencoder}

\end{document}